\documentclass[useAMS,usenatbib]{mn2e}
\usepackage{listings}
\usepackage{graphics}
\usepackage{subfigure}
\usepackage{graphicx}
\usepackage{amssymb}
\usepackage{hyperref,times,cleveref}
\usepackage{setspace,xspace}
\usepackage{multirow}
\usepackage{color}

\newcommand{\rockstar}{\textsc{Rockstar}}

\newcommand{\kms}{km\,s$^{-1}$}

\newcommand{\msun}{$M_{\odot}$}

\newcommand{\mpc}{\,$h^{-1}$Mpc\xspace}
\newcommand{\gpc}{\,$h^{-1}$Gpc\xspace}

\newcommand{\sort}{{\sc sort}\xspace}

\title[Stochastic Order Redshift Technique (SORT)]{Stochastic Order
Redshift Technique (SORT): a simple, efficient and robust method to
improve cosmological redshift measurements}

\author[Tejos, Rodr\'iguez-Puebla, Primack]{
\parbox[t]{\textwidth}{
\vspace{-1.0cm}
Nicolas Tejos,$^{1}$\thanks{E-mail: nicolas.tejos@pucv.cl}
Aldo Rodr\'iguez-Puebla$^{2}$ and Joel R. Primack$^{3}$
}
\vspace*{6pt} \\
$^1$ Instituto de F\'isica, Pontificia Universidad Cat\'olica de Valpara\'iso,
Casilla 4059, Valpara\'iso, Chile\\
$^2$ Instituto de Astronom\'ia, Universidad Nacional Aut\'onoma de M\'exico, A. P. 70-264, 04510, M\'exico, D.F., M\'exico\\
$^3$ Physics Department, University of California, Santa Cruz, CA 95064, USA\\
\vspace*{-0.5cm}}

\begin{document}
\date{MNRAS accepted}

\pagerange{\pageref{firstpage}--\pageref{lastpage}} \pubyear{2017}

\maketitle

\label{firstpage}

\begin{abstract}

We present a simple, efficient and robust approach to improve
cosmological redshift measurements. The method is based on the
presence of a reference sample for which a precise redshift
number distribution ($dN/dz$) can be obtained for different
pencil-beam-like sub-volumes within the original survey. For each
sub-volume we then impose: (i) that the redshift number
distribution of the uncertain redshift measurements matches the reference
$dN/dz$ corrected by their selection functions; and (ii) the rank
order in redshift of the original ensemble of uncertain
measurements is preserved. The latter step is motivated by the
fact that random variables drawn from Gaussian probability
density functions (PDFs) of different means and arbitrarily large
standard deviations satisfy stochastic ordering. We then repeat
this simple algorithm for multiple arbitrary pencil-beam-like
overlapping sub-volumes; in this manner, each uncertain
measurement has multiple (non-independent) ``recovered''
redshifts which can be used to estimate a new redshift PDF. We
refer to this method as the Stochastic Order Redshift Technique
(\sort). We have used a state-of-the art $N$-body simulation to
test the performance of \sort under simple assumptions and found
that it can improve the quality of cosmological redshifts in an
robust and efficient manner. Particularly, \sort redshifts ($z_{\rm sort}$)
are able to recover the distinctive features of the so-called
`cosmic web' and can provide unbiased measurement of the
two-point correlation function on scales $\gtrsim 4$\mpc. Given
its simplicity, we envision that a method like \sort can be
incorporated into more sophisticated algorithms aimed to exploit
the full potential of large extragalactic photometric surveys.

\end{abstract}


\section{Introduction}\label{sec:intro}

Observational constraints of the galaxy distribution are of
fundamental importance for cosmology and astrophysics. As tracers
of the underlying matter distribution, the actual
three-dimensional positions of galaxies contain relevant
information regarding the initial conditions of the Universe,
cosmological parameters, and the nature of dark matter and dark
energy \citep[e.g.][]{Plionis2002, Cole2005, Li2011,
Bos2012, Gillet2015, Cai2015}. This so-called `cosmic web' also affects the
physical condition of the bulk of baryonic matter residing in the
intergalactic medium \citep[e.g.][]{Cen1999, Dave2001,Shull2012},
which in turn could also shape the evolution of galaxies themselves
\citep[e.g.][]{Mo2005, Peng2010, Lu2015, Peng2015, Aragon-Calvo2016}.

Of particular interest is to resolve the cosmic web {and the
density distribution} on scales $\lesssim 1-10$\mpc, where the
clustering power of matter is larger. At these scales, galaxies
form an intricate network of filaments, sheets and nodes, while
also leaving vast volumes virtually devoid of luminous matter
(i.e. galaxy voids).\footnote{But note that such galaxy voids can
still contain baryonic matter in the form of highly ionized
hydrogen \citep[e.g.][]{Penton2002, Tejos2012}.} In order to
access the rich information provided by these complex patterns,
one must survey galaxies with redshift precision comparable or
smaller than such scales.

State-of-the-art photometric redshifts can achieve redshift
precisions of $\sigma_z^{\rm ph} \approx 0.02$ \citep[e.g. for
SDSS at $z<0.6$,][and references therein]{Beck2016}, which
correspond to scales of $\sim 70$\mpc at $z\sim 0.5$. This is
essentially about an order of magnitude larger than the precision
needed to resolve most of the cosmic web patterns. This precision
is unlikely to improve drastically over the next decade; the
photometric redshift technique usually relies on fitting spectral
energy distributions (SEDs) using a fixed set of relatively broad
band filters. Thus, despite having access to deeper observations,
more realistic SED templates, better characterization of
systematics, and/or more complex fitting algorithms, it is the
widths of the photometric filters that determine the intrinsic
limit precision of such techniques. Even photometric surveys that
use a comparatively large number of contiguous medium band
filters \citep[e.g. ALHAMBRA;][]{Arnalte-Mur2014} do not seem to
significantly improve over a $\sigma_z^{\rm ph} \approx 0.02$.

In such an scenario, it seems inevitable to rely (at least
partially) on precise spectroscopic redshifts. As opposed to
broad band filters, the spectroscopic redshift estimations come
from resolving spectral features (usually narrow; e.g.
emission/absorption lines, continuum breaks, etc.), providing
redshift precisions mostly limited by the resolution of the
spectrograph. Even low-resolution ($R\equiv \frac{\lambda}{\Delta
\lambda}\approx 200$) spectroscopy can easily reach $\sim 1$\mpc
scale precision. However, spectroscopic redshifts are
comparatively more expensive due to the combination of reduced
sensitivity, spectral coverage, and field-of-view coverage.

Due to these limitations, it is sensible to adopt an hybrid
approach, where the precise redshift estimations (e.g.
spectroscopic) of a small subsample of the survey are used to
{\it improve} the less precise cosmological redshift measurements
(e.g. photometric) of the majority of the targets in the survey.
This is in principle possible assuming that most of the luminous
galaxies\footnote{Or other biased tracers of the underlying
matter distribution.} occupy relatively small volumes. In such a
case, small reference samples could still provide a sensible
mapping of large-scale structures which can be used to better
constrain the positions along the line-of-sight of relatively
close neighbours. 

The idea of using spectroscopic samples to empirically improve
the redshift distribution of photometric samples is not new
\citep[e.g.,][]{Padmanabhan2005,Sheth2007,Lima2008,Cunha2009}. In
general, these studies focus on improving the underlying redshift
distribution of a photometric sample rather than individual
galaxy redshifts; thus, those methods are appropriate for
applications where the individual information of galaxies is not
required. Some examples of these include surveys aimed at
studying Baryonic Acoustic Oscillations (BAOs), weak lensing,
growth of density perturbations, dark energy from galaxy
distributions and the integrated Sachs-Wolf effect.

Alternative methods exist for improving the individual
photometric redshift of galaxies. For example, the method
developed in \citet{Aragon-Calvo2015} uses the information of the
cosmic web itself to find the maximum likelihood locations of
individual photometric galaxies to the nearest cosmic web
`element'. Another approach is to use the spatial clustering of
galaxies \citep[e.g.][]{Phillipps1987, Landy1996, McQuinn2013,
Menard2013}. These methods use the information of both the
redshift distribution and angular clustering of a spectroscopic
reference sample to improve the individual redshifts of a
photometric sample. Particularly, the implementation presented by
\citet{Menard2013} has been recently successfully applied to the
the Two-Micron All-Sky Survey (2MASS) and the Sloan Digital Sky
Server (SDSS) \citep{Rahman2016a, Rahman2016b}. The virtue of
these methods is that they can provide a redshift
measurement that is essentially {\it independent} of the
photometric redshift based on SEDs, thus both probability density
distributions (PDFs) can be multiplied to infer an improved
redshift measurement for individual sources. Unfortunately, none
of these methods are easily accessible to the community and also
these tend to be computationally expensive. Another limitation is
that they usually rely on wide field surveys and it is not
straightforward how to apply them to much narrower geometries
(particularly pencil-beam-like ones).

This paper presents a very simple and complementary method to
improve cosmological redshift measurements obtained on surveys
with arbitrary geometries. Broadly speaking, the method considers
a patch on the sky where initially two kinds of redshift
measurement exist: uncertain (e.g. photometric) and precise (e.g.
spectroscopic). The ones with precise redshifts will be used as a
`reference sample' and it is of course is expected that these
correspond to a small fraction of the total number of objects. We
then rank order both distributions in order to find a correlation
between the reference and the uncertain sample. As we will show
later, this idea is motivated by the fact that random variables
drawn from Gaussian PDFs of different means and arbitrarily large
standard deviations satisfy stochastic ordering. We thus refer to
this method as the Stochastic Order Redshift Technique (\sort).

By construction, \sort is non-parametric as it does not need to
assume any functional form for either the distribution of
redshift or the relationship between spectroscopic and
photometric redshifts. Thus, the power of \sort relies on its
simplicity and versatility. We show that \sort is indeed robust
and that it can provide unbiased measurement of the
redshift-space two-point correlation function on scales $\gtrsim
4$\mpc, while also recovering the distinctive features of the
cosmic web (voids, filaments, clusters). Given its simplicity, we
expect that a method like \sort could be incorporated into more
sophisticated algorithms aimed to improve cosmological redshifts
in order to fully exploit the potential of large extragalactic
photometric surveys like the Dark Energy Survey
\citep[DES;][]{Flaugher2012, DarkEnergySurveyCollaboration2016}
or the The Large Synoptic Survey Telescope
\citep{LSSTScienceCollaboration2009}. Moreover, because \sort can
be applied to both wide and narrow surveys, we also expect to be
suitable for current surveys like COSMOS \citep{Scoville2007,
Koekemoer2007} or CANDELS \citep[][]{Grogin2011, Koekemoer2011}.

This paper is organized as follows. In \Cref{sec:method}, we
describe the method, while in \Cref{sec:mock} we  describe the
galaxy survey used to study its performance. In
\Cref{sec:results} we present the results of applying \sort to
our mock galaxy survey, including redshift PDFs, two-point
correlation functions and inferred three-dimensional
distributions of galaxies. In \Cref{sec:discussion} we provide a
discussion regarding information content, efficiency,
versatility, limitations, and a comparison to other methods.
\Cref{sec:summary} presents a summary and main conclusions.

\section{The method}\label{sec:method}

We propose a simple and efficient method to improve cosmological
redshift measurements whose uncertainties are larger
than that of the intrinsic `cosmic web' scales. The general idea
behind the method is based on the use of a reference sample and
{\it stochastic order} as we describe below.

\subsection{General considerations}\label{sec:general}

    \begin{figure*}
    
    \begin{minipage}{1\textwidth}
    \centering
    \includegraphics[width=1\textwidth]{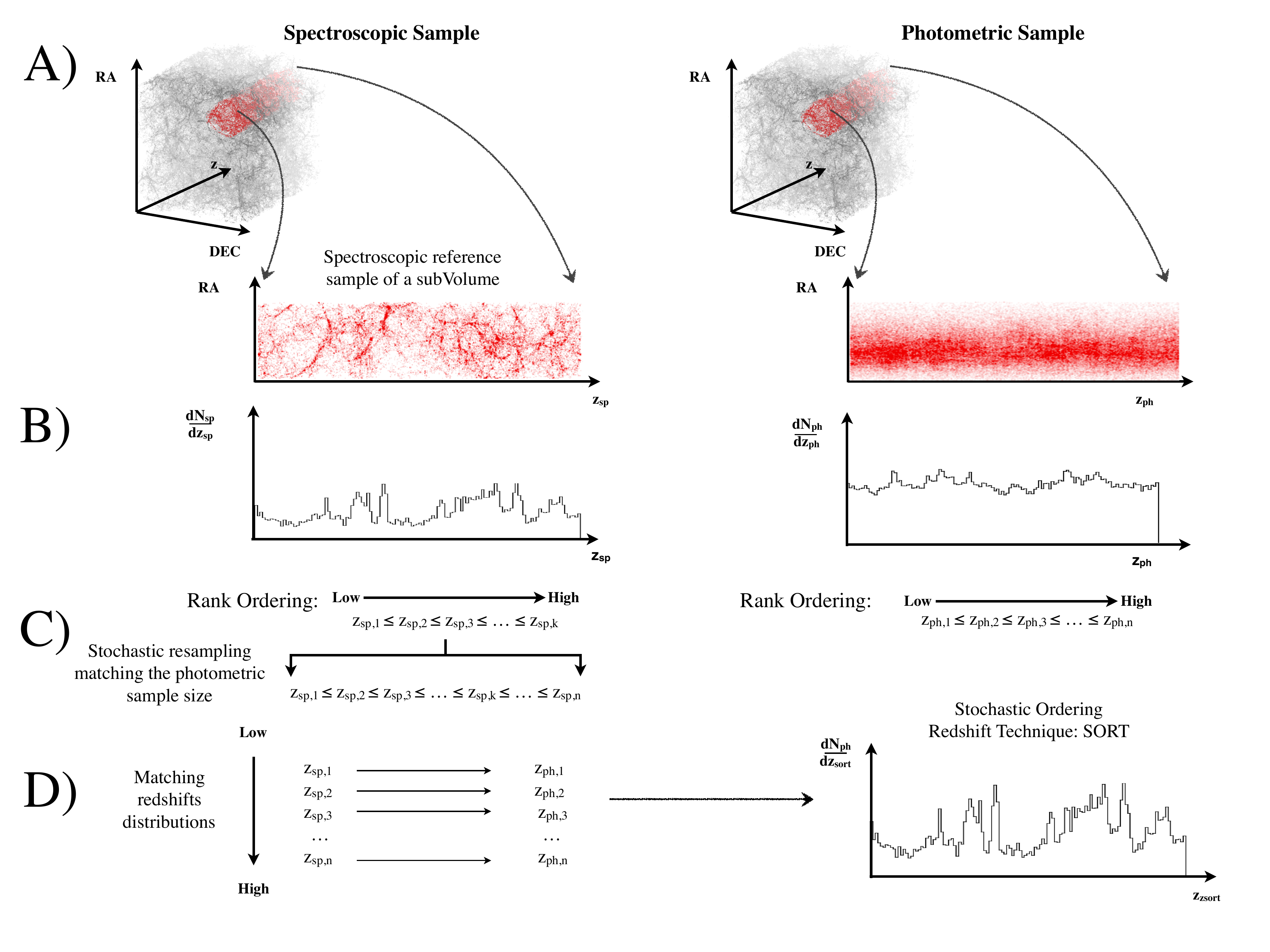}
    \end{minipage}

    \caption{An illustration of our method based on stochastic
order. {\it Step (A):} In a given pencil-beam-like sub-volume we
define two samples of extragalactic objects depending on the
accuracy of their cosmological redshift determination: precise
(e.g. spectroscopic; left panels) and uncertain (e.g.
photometric; {right panels}). {\it Step (B):} From both samples
we observe a redshift number density, $dN/dz$. In principle, the
uncertain $dN/dz$ (left panel) is a {\it nosier} version of the
precise one (right panel). {\it Step (C):} From the precise
distribution we create a new re-sampled redshift distribution
matching the number of objects in the uncertain sample (left
panel) and sort it from low to high redshift. We also sort the
observed uncertain redshift distribution from low to high
redshift (right panel). {\it Step (D):} Finally, we perform a
one-to-one match between the recovered distribution in the left
and right panels of step (C). We refer to this simple algorithm
as Stochastic Ordering Redshift Technique (\sort).}

\label{fig:diagram}

\end{figure*}

Let us consider a survey of $N_{\rm tot}$ galaxies in a volume
$V$. Suppose that in this galaxy survey we find that for a number
of $N_{\rm ph}$ galaxies the quality of their individual
redshifts is poor (hereafter referred to as the photometric
sample), and that $N_{\rm sp}$ galaxies have a reliable redshift
estimation (referred to as the spectroscopic
sample).\footnote{But note that there could be reliable redshift
estimations from photometry, and/or poor estimations from
spectroscopy (e.g. objects with lack of narrow spectral
features).}  The upper row of \Cref{fig:diagram} illustrates the
situation described above. It is evident that the probability
redshift distributions of these two subsamples are different:
while the spectroscopic sample traces a complex distribution, such
information is somewhat lost in the photometric sample (see
B row of \Cref{fig:diagram}). Observationally, one can
constrain these probability distributions, $P(z)$, by means of
their redshift number density, $dN/dz(z)$, i.e. the number of
galaxies per unit redshift in a given sample. Thus,

\begin{equation} 
P_{\rm ph}(z_{\rm ph}) = \frac{1}{S_{\rm ph}} \frac{1}{N_{\rm ph}}\frac{dN_{\rm ph}}{dz_{\rm ph}}(z_{\rm ph}) 
\label{eq:dndz_ph}
\end{equation}

\noindent and,

\begin{equation} 
P_{\rm sp}(z_{\rm sp}) = \frac{1}{S_{\rm sp}} \frac{1}{N_{\rm sp}}\frac{dN_{\rm sp}}{dz_{\rm sp}}(z_{\rm sp}) 
\label{eq:dndz_sp}
\end{equation}

\noindent where $S_{\rm ph}$ and $S_{\rm sp}$ are the selection
functions for these two samples, respectively. The selection
function should depend on redshift, but it can also depend on
position on the sky and any other property of the extragalactic
object (e.g. luminosity, morphology, star-formation rate, etc.).
Assuming that there is a {\it real} redshift probability
distribution traced by galaxies, these distributions can also be
written as

\begin{equation} 
P_{\rm ph}(z_{\rm ph}) = \int  \mathcal{R}_{\rm ph}(z-z_{\rm ph}) P_{\rm real}(z)\, dz
\label{eq:p_ph}
\end{equation}

\noindent and, 

\begin{equation} 
P_{\rm sp}(z_{\rm sp}) = \int \mathcal{R}_{\rm sp}(z-z_{\rm sp}) P_{\rm real}(z)\, dz
\label{eq:p_sp}
\end{equation}

\noindent where $\mathcal{R}_{\rm ph}$ and $\mathcal{R}_{\rm sp}$
are the redshift uncertainties of the photometric and
spectroscopic samples, respectively. The redshift {\it
resolution} of these estimations are therefore limited by the
redshift uncertainties of the galaxies in their respective
samples (see panel B of \Cref{fig:diagram} for an illustration);
one can consider $P_{\rm ph}$ be a {\it noisier} version of
$P_{\rm sp}$.

Assuming that the subsample of $N_{\rm sp}$ galaxies with
spectroscopic redshifts is `statistically relevant'---i.e. that
they accurately trace the underlying cosmic structures such as
clusters, filaments, walls and voids present in the volume $V$---
this naturally suggests that the spectroscopic subsample
can be used to improve the quality of the redshift distribution
of the photometric sample. In other words, we can assume that the
probability redshift distribution of the photometric sample,
$P_{\rm ph}$,  is the probability redshift distribution of the
spectroscopic sample, $P_{\rm sp}$, convolved with an unknown
kernel $\mathcal{G}$,

\begin{equation} 
P_{\rm ph}(z_{\rm ph}) =  \int \mathcal{G}(z_{\rm ph} - z_{\rm sp}) P_{\rm sp}(z_{\rm sp}) dz_{\rm sp}
\label{eq:G_def}
\end{equation}

\noindent Therefore, the problem of improving photometric
redshift estimations may reduce to constrain all the moments of
$\mathcal{G}$, particularly its mean relation (i.e. $z_{\rm ph} -
z_{\rm sp}$) and the dispersion around the mean. The general
approach has been then to constrain $\mathcal{G}$ and use
\Cref{eq:G_def} to improve the redshift estimations of the
photometric sample.\\

In this paper we propose a complementary approach, in which we do
not constrain $\mathcal{G}$ itself to obtain the underlying
$P(z)$, but use the overall relation between their cumulative
distributions instead. Let us consider the simple case when the
correlation between $z_{\rm sp}$ and $z_{\rm ph}$ is a one-to-one
monotonic relation with zero scatter, i.e. $\mathcal{G}(z_{\rm
ph} - z_{\rm sp})$ becomes the Dirac delta function. Then 
\Cref{eq:G_def} can be written as:

\begin{equation}
    \frac {dN_{\rm ph}} {dz_{\rm ph}} (z_{\rm ph})=  
    \frac{N_{\rm ph}}{N_{\rm sp}} \frac {S_{\rm ph}}{S_{\rm sp}}\frac {dN_{\rm sp}} {dz_{\rm sp}}(z_{\rm sp}(z_{\rm ph})) \frac {dz_{\rm sp}} {dz_{\rm ph}},
	\label{eq:dif_matching}
\end{equation}

\noindent where we use the definitions of \Cref{eq:dndz_ph,eq:dndz_sp}. 
The above Equation can be written in terms of the following integrals:
\begin{equation} 
    \int_{z_{\rm ph}}^{\infty}\frac {dN_{\rm ph}}{dz'_{\rm ph}} dz'_{\rm ph} = 
    \frac{N_{\rm ph}}{N_{\rm sp}} \int_{z_{\rm sp}}^{\infty} 
    \frac{S_{\rm ph}}{S_{\rm sp}}\frac {dN_{\rm sp}} {dz'_{\rm sp}} dz'_{\rm sp}. 
    \label{eq:matching_z} 
\end{equation} 

An easy way to solve \Cref{eq:matching_z} is just rank ordering
spectroscopic galaxies by their redshift and assigning them to
photometric galaxies also ranked by redshift (see bottom panel of
\Cref{fig:diagram} and \Cref{eq:rec_samp} below). Obviously, in
the presence of scatter in the relation $\mathcal{G}$ this
solution is not strictly valid but just an approximation. In this
paper, we explore how good such an approximation is, and show
that it is indeed suitable for improving photometric redshifts.

\subsection{Stochastic Order Redshift Technique (SORT)}\label{sec:sort}

We solve \Cref{eq:matching_z} by sorting the $N_{\rm ph}$ {\it
observed} photometric redshifts such that $z_1^{\rm obs} \le
z_2^{\rm obs} \le \dots \le z_{N_{\rm ph}}^{\rm obs}$, and assign
them $N_{\rm ph}$ sorted {\it recovered} redshifts, randomly
sampled from,

\begin{equation} 
\frac{N_{\rm ph}}{N_{\rm sp}} \frac{S_{\rm ph}}{S_{\rm sp}} \frac{dN_{\rm sp}}{dz}(z) \to \{z_1^{\rm rec}, z_2^{\rm rec},\dots,z_{N_{\rm ph}}^{\rm rec}\}
\label{eq:rec_samp}
\end{equation}

\noindent such that $z_1^{\rm rec} \le z_2^{\rm rec} \le \dots
\le z_{N_{\rm ph}}^{\rm rec}$. This provides a straightforward
one-to-one mapping between the observed and {\it recovered}
photometric redshift distributions as $z_i^{\rm obs}
\leftrightarrow z_i^{\rm rec}$, for $i \in \{1,2,\dots,N_{\rm
ph}\}$ (see bottom panel of \Cref{fig:diagram}). This is a
simplistic but powerful approach, particularly because
photometric samples are expected to satisfy {\it stochastic
ordering}.

We define {\it stochastic order} as follows. Consider two random
variables, $X_i$ and $X_j$, being drawn from two arbitrary
probability density functions (PDFs), $P_i$ and $P_j$ in the
domain $x$. Then, we say that $X_j$ is stochastically greater
than or equal to $X_i$ if and only if their PDFs satisfy,

\begin{equation} 
P_i(X_i > x) \le P_j(X_j > x) \ \ \  \forall x \ \ \rm{,}
\label{eq:order}
\end{equation}

\noindent \citep[e.g.][]{Shaked2007}. This is equivalent to saying
that,

\begin{equation} 
\int_{x}^{\infty} P_i(x')\,dx' \le \int_{x}^{\infty} P_j(x')\,dx' \ \ \ \forall x \ \ \rm{.}
\label{eq:order2}
\end{equation}

Let us now consider the case of individual redshift estimations,
$z_i$ and $z_j$, whose PDFs are given by Gaussians having the
same (arbitrarily large) standard deviations, centred at $z_i$
and $z_j$, respectively, and satisfying $z_i < z_j$. If we treat
these redshift measurements as random variables, it is
straightforward to show that stochastic order is satisfied
(\Cref{eq:order} or \Cref{eq:order2}). Hence, even though their
PDFs may overlap in redshift, the most likely outcome is having
$z^{\rm true}_i \le z^{\rm true}_j$, where $z^{\rm
true}_{\{i,j\}}$ are their true underlying redshifts,
respectively.

Stochastic order ensures transitivity, meaning that independently
of the individual photometric redshift uncertainties, their
observed rank order within the sample most likely matches that of
the underlying true values. Thus, by solving \Cref{eq:matching_z}
using \Cref{eq:rec_samp} this information is also preserved.

Although there may be cases where \Cref{eq:order} does not apply
(e.g. complex PDFs with asymmetries or multiple local maxima), we
can expect that state-of-the-art photometric redshift estimations
provide well behaved PDFs, where stochastic order is satisfied
for a significant fraction of the photometric sample. Under this
assumption, we thus refer to this matching scheme as the
Stochastic Order Redshift Technique (\sort).\\

We emphasize that \sort is a statistical method and should be only
applied to ensembles rather than individual measurements. We also
note that this simplistic version of \sort could eventually lead
to catastrophic redshifts, i.e. $z_i^{\rm rec}$ that are not
statistically consistent with the $z_i^{\rm obs}$ individual
PDFs. Indeed, this is  the case because in its current
implementation we are not using the information provided by the
individual photometric uncertainties. We chose this
approach so we can explore the advantages/disadvantages of the
method in its simplest form (but see \Cref{sec:information} for
how to easily prevent catastrophic redshift assignments).

In the following we test \sort using a mock galaxy survey drawn
from a state-of-the-art $N$-body cosmological simulation aimed at
reproducing the low-$z$ ($z\lesssim 0.3$) SDSS. We emphasize that
even though we will mainly refer to a galaxy survey, in practice
we can apply this method to any kind of luminous extragalactic
object.

\section{Simulation}\label{sec:mock}

In order to test the performance of \sort we create a
magnitude-limited mock galaxy survey on the $N$-body
MultiDark-Planck simulation \citep{Klypin2016}. The
MultiDark-Planck simulation is a very high resolution simulation
based on a $\Lambda$CDM cosmology with the  results from the
Planck Collaboration \citep{PlanckCollaboration2016}: $\Omega_{\rm M} = 0.307$,
$\Omega_{\Lambda} = 1 - \Omega_{\rm M}$, $\Omega_{\rm B} =
0.048$, $\sigma_8 = 0.829$, $n_s=0.96$ and $h = 0.678$. The
length  of the simulation is $1$\gpc per side and it contains
$3840^3$ particles of mass $m_{\rm p} = 1.5 \times 10^{9}$\,
$h^{-1}M_{\odot}$.

\subsection{Mock galaxy survey}

We created a mock magnitude limited galaxy survey based on the MultiDark
Planck simulation \citep{Klypin2016, Rodriguez-Puebla2016}.\footnote{The 
outputs of the simulation are available in \url{http://hipacc.ucsc.edu/Bolshoi/MergerTrees.html}.}  Dark
matter haloes and subhalos were identified using the \rockstar\
phase-space temporal halo finder \citep{Behroozi2013}. Halo mass
is defined using the virial overdensity given by the spherical
collapse model which corresponds to $\Delta_{\rm} = 333$ at $z =
0$ for our adopted cosmology. We then imposed a minimum halo mass of
$10^{12}h^{-1}$\msun, and assume that each one of those
haloes/subhaloes will host a luminous galaxy. Galaxies were
assigned via abundance matching \citep{Vale2004}  by using the
luminosity function of the SDSS DR7 \citep{Yang2009} in the
$r$-band that will be denoted simply as $M$.

We placed a virtual observer at one of the corners of the
simulated box and projected each galaxy position in co-moving
Cartesian coordinates into Spherical Coordinates assuming the
pole to be aligned with one of the Cartesian axes. In this
manner, the simulated box is projected into a single spherical
octant. We assign the azimuthal angle to the so-called right
ascension of the Celestial Coordinates (RA), and the complement
of the polar angle to the so-called declination of the Celestial
Coordinates (Dec.). We used the total  co-moving distance of each
galaxy to the position of the virtual observer to define
cosmological redshifts, $z_{\rm cos}$, from our adopted
cosmology.

Apparent magnitudes, $m$, were obtained from

\begin{equation}
m = M + \mu(z_{\rm cos})
\label{eq:magnitude}
\end{equation}

\noindent where $\mu(z)$ is the distance modulus as a function of
redshift. We imposed a maximum apparent magnitude limit in the
$r$-band of $m^{\rm max}_{r} = 18.5$\,mag. This criterion led to
$127\,993$ galaxies, all  at cosmological distances smaller
than the size of the simulation box.

We define `true' redshift, $z_{\rm true}$, from

\begin{equation}
1+ z_{\rm true} = (1+z_{\rm cos})(1+\frac{\Delta v_{\rm los}}{c}) \ \rm{,}
\end{equation}

\noindent where $\Delta v_{\rm los}$ is the simulated velocity
component along the line-of-sight (i.e. peculiar velocity), and
$c$ is the speed of light. This $z_{\rm true}$ will be considered
the intrinsic true value of an individual galaxy's redshift. In
order to simulate an observed redshift, we added noise in the
form of

\begin{equation}
z_{\rm obs} = z_{\rm true} + \delta_z (1+z_{\rm true})\ \rm{,}
\label{eq:zobs}
\end{equation}

\noindent where $\delta_z$ is a value randomly sampled from a Gaussian
distribution centred at zero with standard deviation $\sigma_z$.

\subsection{Galaxy subsamples} 

We define two main subsamples of galaxies based on two different level of
noise applied to their $z_{\rm true}$ (see \Cref{eq:zobs}),

\begin{itemize}

\item {\it The photometric sample:} defined by using a $\sigma_z^{\rm
ph} = 0.02$ encompassing $70\%$ of the magnitude limited sample
(randomly chosen).\smallskip

\item {\it The spectroscopic sample:} defined by using $\sigma_z^{\rm
sp} = 0.0001$ for the remaining $30\%$ of the sample.

\end{itemize}

We chose $\sigma_z^{\rm ph} = 0.02$ as a representative
state-of-the-art photometric precision (this is also comparable to the
goal of the future Large Synoptic Survey Telescope, LSST). The $30\%$
reference fraction was chosen as representative of current large
extragalactic surveys at our chosen magnitude limit (e.g. SDSS). For
simplicity, in the following  we restrict the analysis to a unique
sample using these fiducial values. For a discussion on the
performance of the method using lower fractions of spectroscopic
galaxies and larger photometric uncertainties we refer the reader to
\Cref{sec:versatility}.

 \begin{figure}
    
    \begin{minipage}{0.48\textwidth}
    \centering
    \includegraphics[width=1\textwidth]{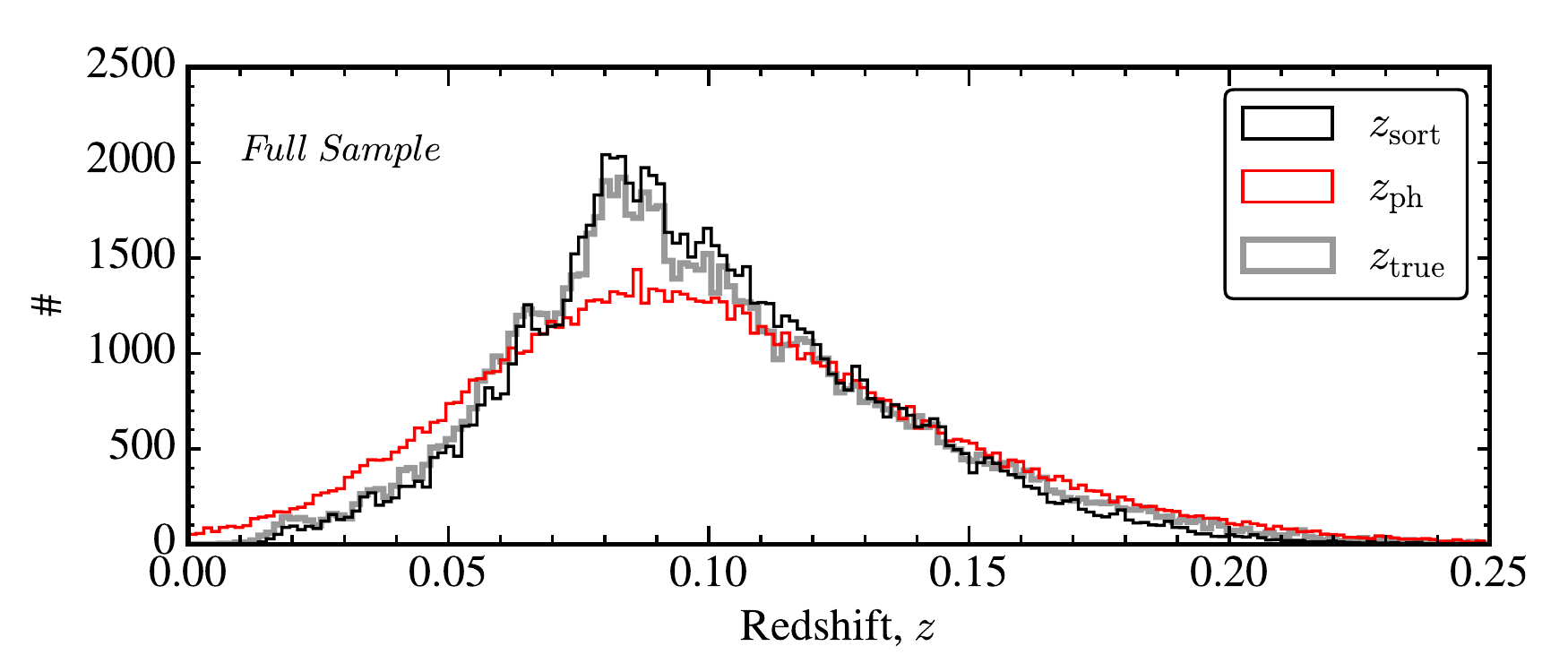}
    \end{minipage}
    \begin{minipage}{0.48\textwidth}
    \centering
    \includegraphics[width=1\textwidth]{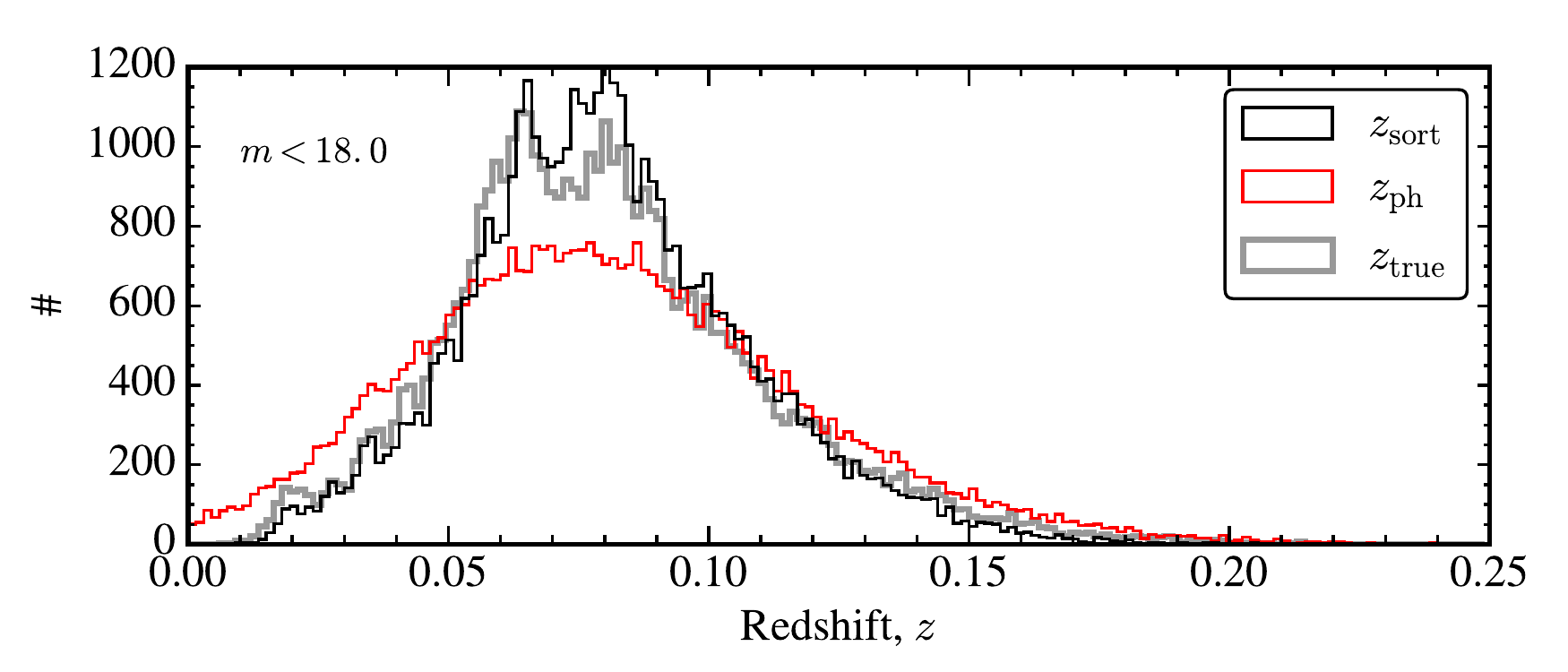}
    \end{minipage}
    \begin{minipage}{0.48\textwidth}
    \centering
    \includegraphics[width=1\textwidth]{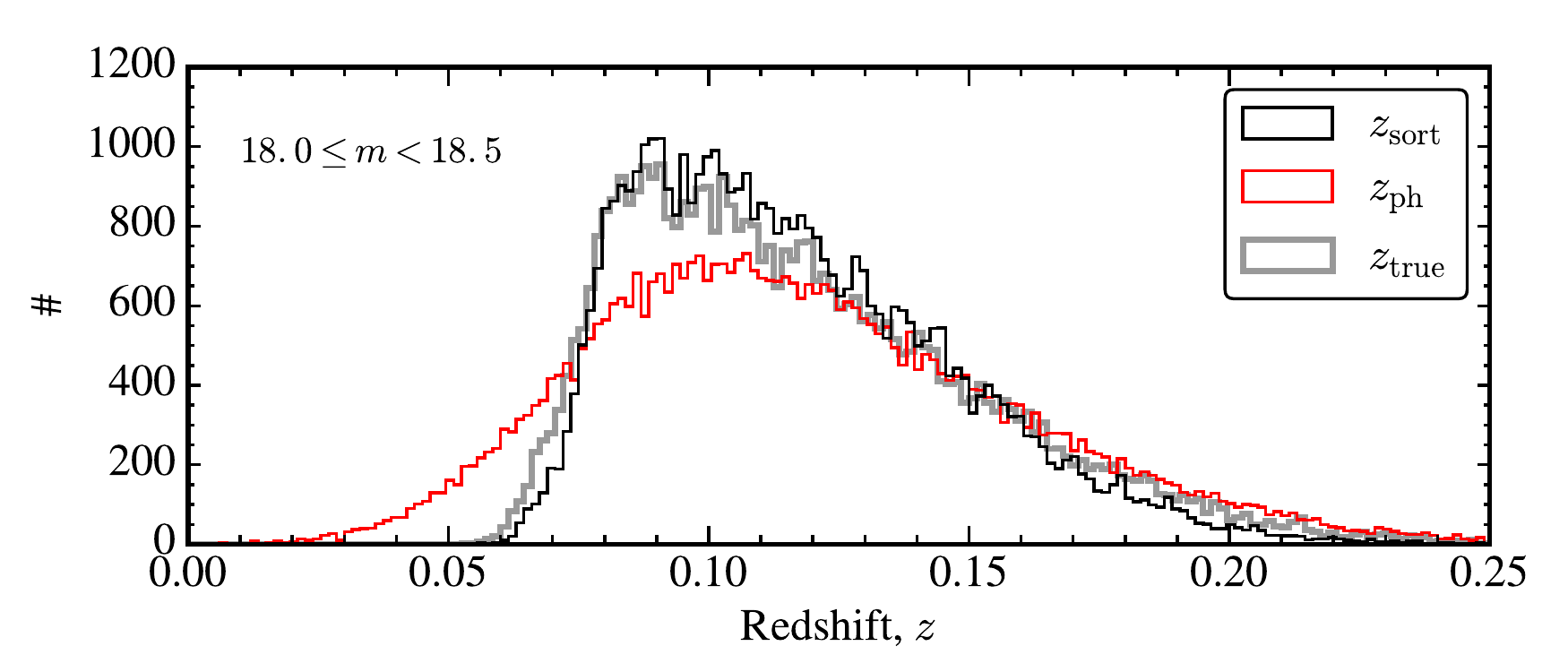}
    \end{minipage}

    \caption{Redshift distributions (arbitrary binning of $0.0015$)
for our photometric mock galaxy survey for different redshifts:
photometric ($z_{\rm ph}$; red histograms), \sort ($z_{\rm sort}$;
black histograms) and true ($z_{\rm true}$; grey histograms). The top
panel shows the full photometric sample, while the middle and bottom
panels split the mock survey to roughly the half-bright and half-faint
subsamples, respectively, by using a $m=18$ limit. See
\Cref{sec:z_dists} for further details.}
\label{fig:z_dists}

\end{figure}



     \begin{figure*}
    \begin{minipage}{0.45\textwidth}
    \centering
    \includegraphics[width=\textwidth]{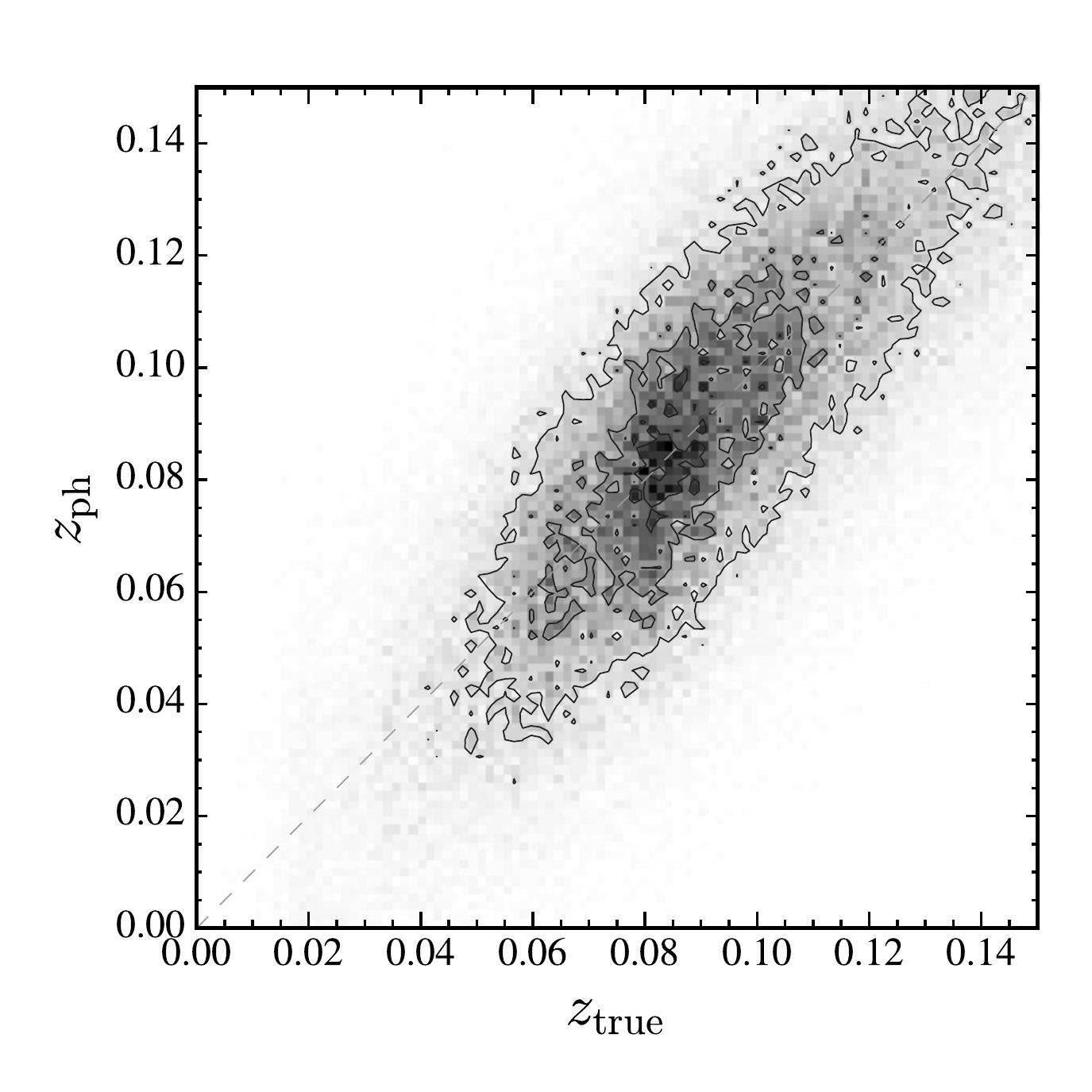}
    \end{minipage}
    \hspace{5ex}
    \begin{minipage}{0.45\textwidth}
    \centering
    \includegraphics[width=\textwidth]{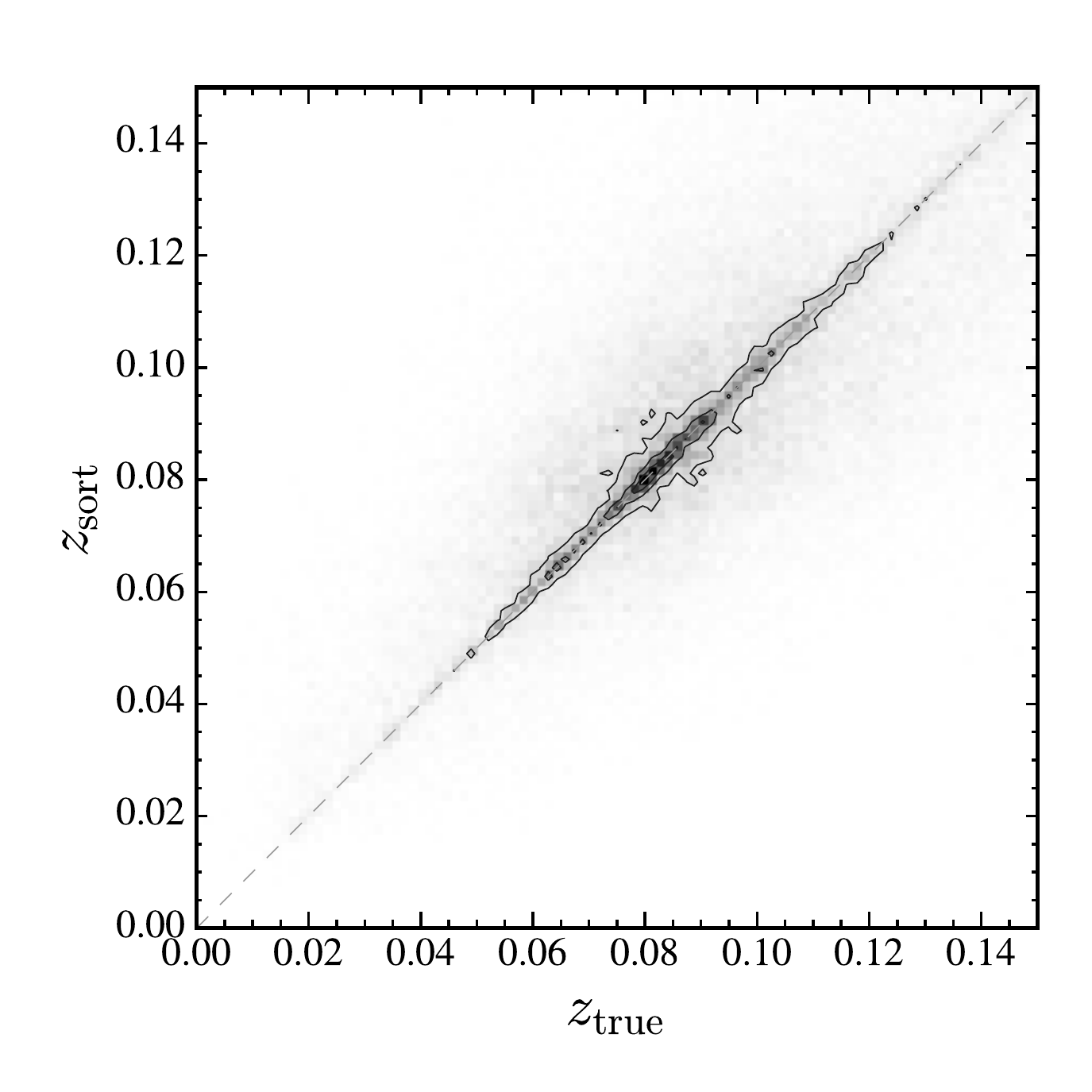}
    \end{minipage}

    \caption{Two-dimensional normalized histograms (arbitrary
binning $0.0015$) for the original photometric redshift ($z_{\rm
ph}$; left panel $y$-axis) and that of \sort ($z_{\rm sort}$;
right panel $y$-axis), both as a function of the underlying true
redshift ($z_{\rm true}$; $x$-axes). The grey scale is linearly
proportional to the number of counts in each normalized bin and
the contours mark the limits where the counts are multiples of
$1/4$ the maximum value. See \Cref{sec:improving} for further details.}

\label{fig:2dhist_comparison}

\end{figure*}

\section{Results}\label{sec:results}

In this section we describe our \sort method applied to the mock galaxy
survey described in the previous section. We will
assume that $30\%$ of our final sample of 127993 galaxies have
spectroscopic redshifts. In reality, this fraction is  a free
parameter however. We refer the reader to Section
\ref{sec:versatility} where we explore the performance of the method
with different choices.


\subsection{Applying the method}\label{sec:apply}

\begin{enumerate}

\item For each individual galaxy with apparent magnitude $m$, we
define a projected area $A$ as a circle of angular radius $R$.

\item We then consider only galaxies in $A$ having apparent magnitudes
within $\Delta m$ from $m$.

\item From these galaxies, we check that at least $N^{\rm ref}_{\rm
min}$ have spectroscopic redshifts. Otherwise, we iterate steps (i) and
(ii) increasing the values of $R$ and $\Delta m$ by $\delta R$ and
$\delta m$, respectively, until the condition is satisfied.

\item We compute a redshift histogram of the $N^{\rm ref}$
galaxies using redshift bins of width $\delta z/3$. We then convolve
this histogram with a Gaussian kernel of standard deviation $\delta
z$, and use the resulting distribution to randomly sample
$N^{\rm rec}_{\rm ph}$ redshifts, where $N^{\rm rec}_{\rm ph}$ is the
number of photometric galaxies within $A$.

\item We apply the stochastic order matching scheme described in
\Cref{sec:sort} (see \Cref{eq:rec_samp}) to obtain {\it recovered} 
redshifts for each individual photometric galaxy.

\end{enumerate}

We repeated this algorithm for {\it all} the photometric galaxies in
the sample. In this manner, each photometric galaxy has a {\it list}
of {recovered} redshifts, each one coming from an independent random
sampling of their respective reference (spectroscopic) sample. This is
equivalent to having {\it adaptive} Monte Carlo realizations, as
galaxies in denser regions will be sampled more times than galaxies in
less dense regions. For simplicity, we finally take the median value
of the aforementioned {recovered} distribution list as the actual
unique {\it recovered} \sort redshift, $z_{\rm sort}$.

For the results of this paper we applied the aforementioned
algorithm using the following parameters: $R=1$\,degree, $\Delta
m = 0.2$\,mag, $\delta R=0.1$\,degree, $\delta m = 0.1$\,mag,
$N^{\rm ref}_{\rm min} = 2$ and $\delta z=0.0003$. By choosing
$N^{\rm ref}_{\rm min} = 2$ we make sure that for each iteration
there are at least $2$ galaxies with a spectroscopic redshift
measurement as reference. Because of our survey is magnitude
limited, for the brightest galaxies this condition means that the
radius of the search ends up being larger than the fiducial value
$R=1$\,degree, up to a factor of $\approx 1.5-2.5\times$ for
those with magnitudes $m\sim 15-14$\,mag respectively. The
fraction of bright galaxies is very small and we do not consider
this issue to be a major limitation of our method; in any case,
the brightest objects are the cheapest to get a spectroscopic
redshift for, hence making it feasible to eventually correct for
this in the future.

We have chosen a relatively small $\Delta m= 0.2$\,mag in order
to ensure a similar selection function for galaxies for the
photometric and spectroscopic samples as a function of $m$. In
this manner we avoid introducing shot noise from correcting for
the different selection functions with a sparse sampling.

    \begin{figure}
    
    \begin{minipage}{0.48\textwidth}
    \centering
    \includegraphics[width=\textwidth]{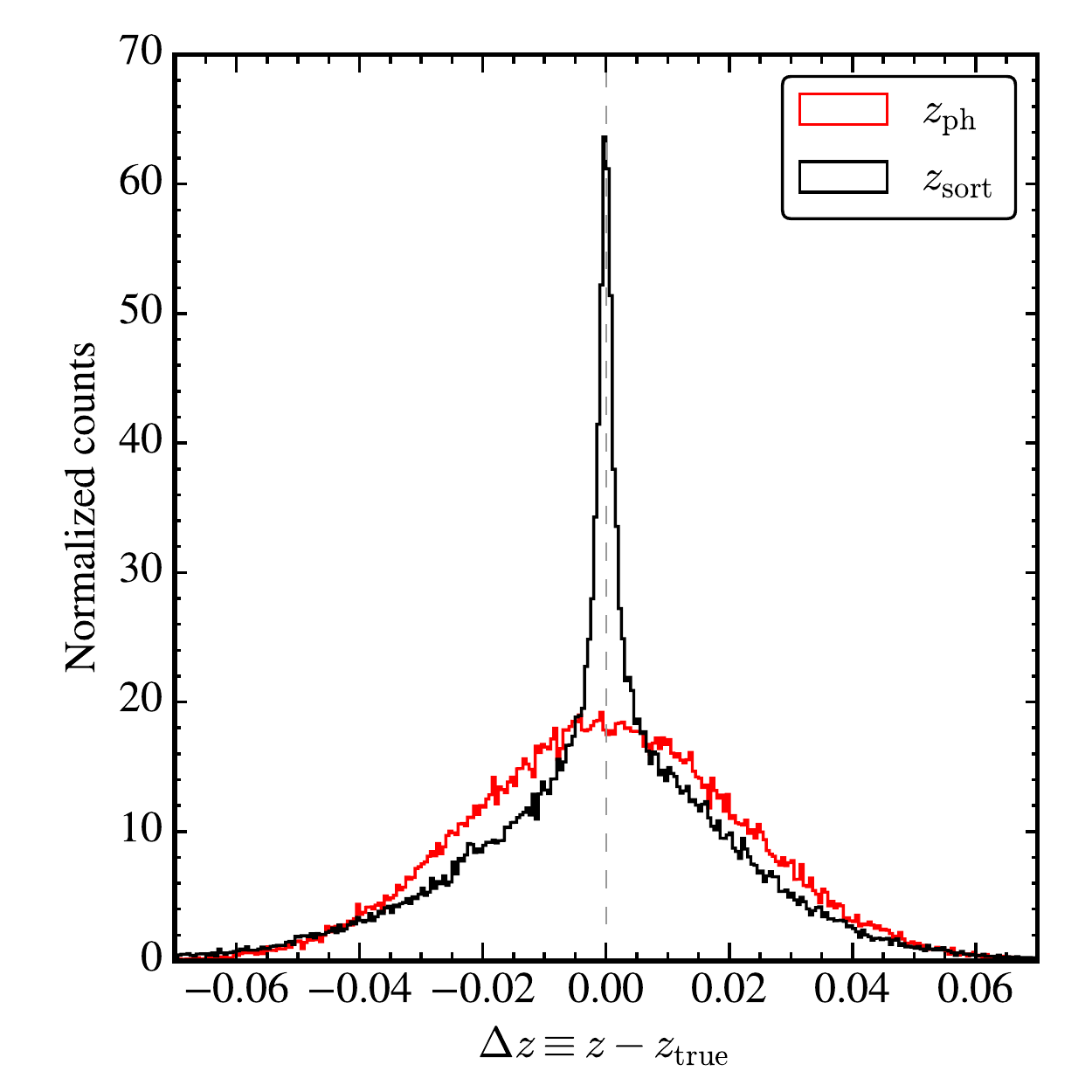}
    \end{minipage}

    \caption{Normalized distribution (arbitrary binning of $0.0005$) of the redshift differences with
respect to the true redshift $z_{\rm true}$, for both the original
photometric redshifts ($z_{\rm ph}$, red histogram), and those from
our \sort method ($z_{\rm sort}$, black histogram). See
\Cref{sec:improving} for further details.}

\label{fig:hist_comparison}

\end{figure}

     \begin{figure*}
    \begin{minipage}{0.45\textwidth}
    \centering
    \includegraphics[width=\textwidth]{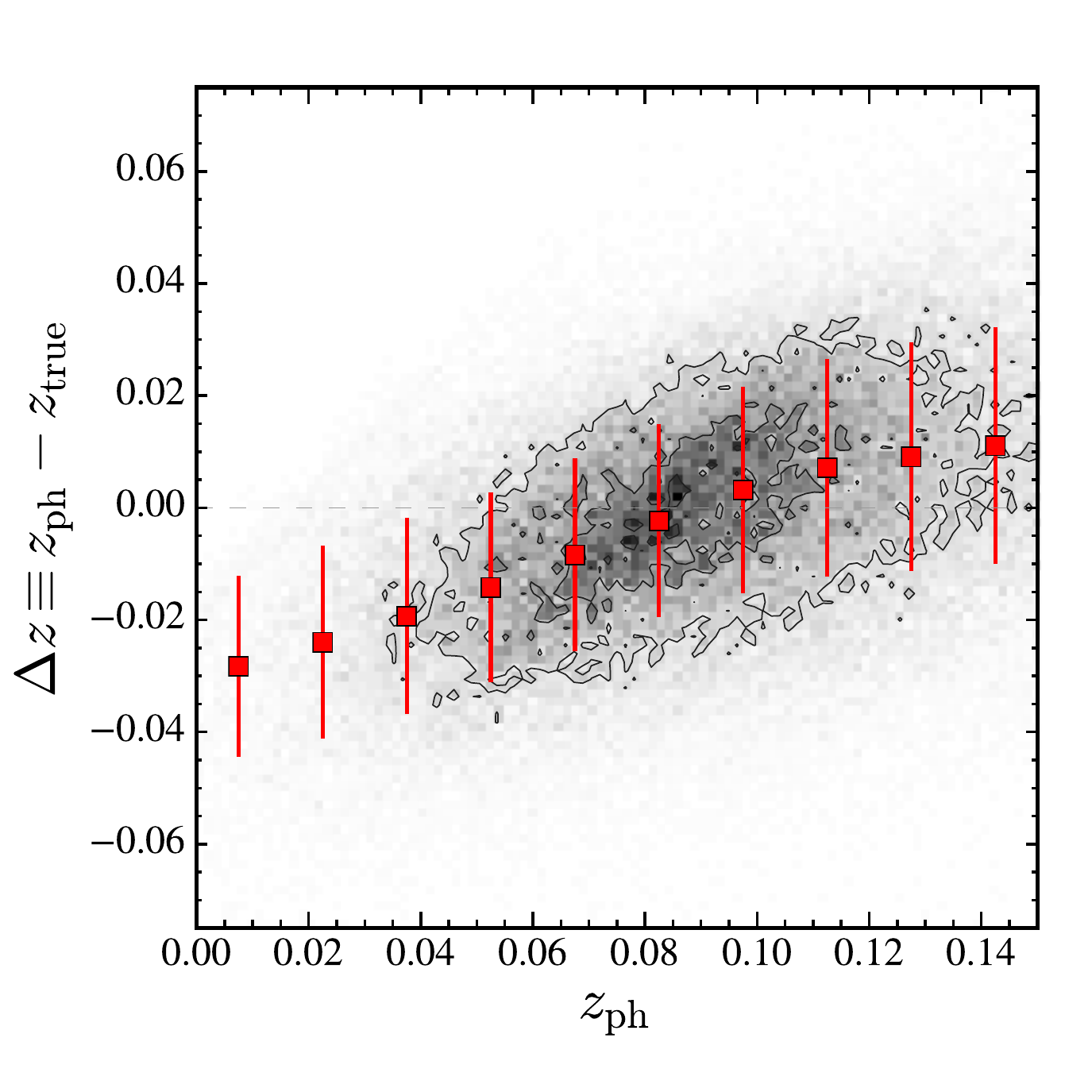}
    \end{minipage}
    \hspace{5ex}
    \begin{minipage}{0.45\textwidth}
    \centering
    \includegraphics[width=\textwidth]{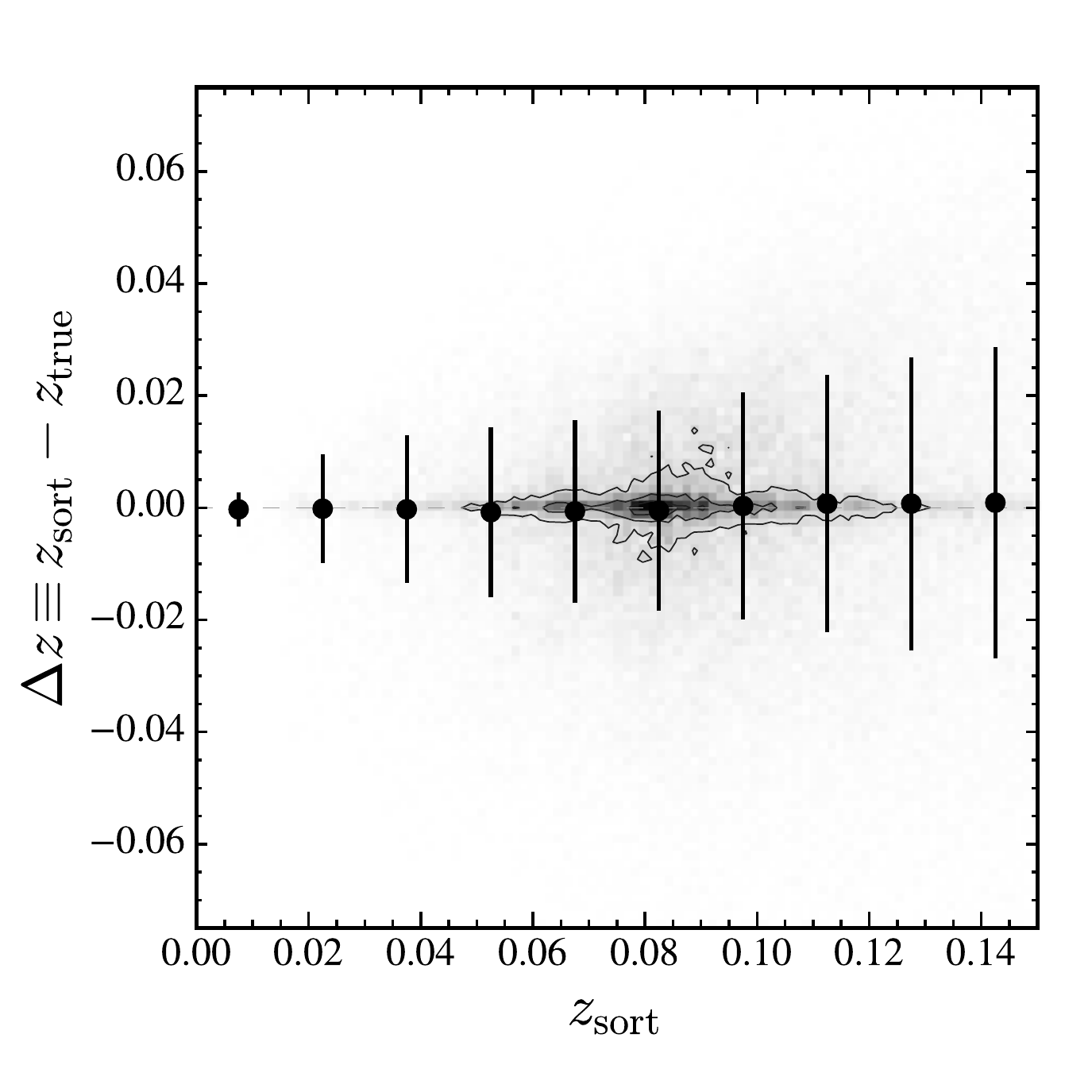}
    \end{minipage}

    \caption{Mean redshift difference with respect to the true
underlying value, $\Delta z \equiv z - z_{\rm true}$ using the
original photometric redshifts ($z=z_{\rm ph}$; left panel, red
squares) and \sort redshifts ($z=z_{\rm sort}$; right panel, black
circles) both as a function of the {\it used} redshift $z$ (arbitrary
binning of $0.015$). We also show the underlying distributions as
normalized two-dimensional histograms with the same binning and contour scheme as
\Cref{fig:2dhist_comparison}. See \Cref{sec:improving} for further
details.}

\label{fig:2dhist_zdiff}

\end{figure*} 

Although the area $A$ can in principle have any arbitrary shape, we
have chosen a circular one only for simplicity. We note that
$1$\,degree corresponds to $\approx 5-10$\mpc at redshifts $z \approx
0.06-0.12$ (where most of our mock galaxies reside). We chose a
$\delta z=0.0003$ which is $3\times$ our adopted $\sigma_z^{\rm
sp}$, corresponding to $\approx 1$\mpc (if cosmological) along the
line-of-sight at these redshifts. Therefore, our chosen $R$ and
$\delta z$ will capture the most relevant scales of the cosmic web
traced by luminous galaxies. We emphasize that our method is very
general, and by construction it only ensures that the {overall}
$P_{\rm gal}(z)$ distributions between the photometric and
spectroscopic samples are statistically consistent with each other
(when corrected for their respective selection functions). However, by
choosing a scale comparable to the cosmic web we also expect to
perform relatively well at recovering the three-dimensional
distribution of galaxies (see \Cref{sec:3d-dist}) and the two-point
auto-correlation function on scales $\gtrsim 5$\mpc (see
\Cref{sec:2pcf}).

\subsection{Redshift distributions}\label{sec:z_dists}

In \Cref{fig:z_dists} we plot the redshift distributions of the mock
galaxies using our different samples: photometric ($z_{\rm ph}$; red
histograms), \sort ($z_{\rm sort}$; black histograms) and true
($z_{\rm true}$; grey histograms). The top panel shows the full
sample, while the middle and bottom panels split the mock survey to
roughly the half-bright and half-faint subsamples, respectively, by
using a $m=18$ limit. As expected, the photometric uncertainties are
large enough to `wash out' most of the large-scale structure
information, resulting in smooth distributions at virtually all
relevant scales. In contrast, the $z_{\rm sort}$ distributions match
the true distributions better than the photometric redshifts, not just
in the overall shape (given mostly by the selection function), but
also the fluctuations induced by the large-scale structure of the
cosmic web itself (i.e. peaks and valleys). This match is not perfect
however, as one can observe that some large-scale structures are
somewhat over/under represented. These discrepancies mainly come from
the sparse sampling of spectroscopic galaxies used as reference at the
low and high ends of the distribution (more evident at $z<0.05$),
which for conservation of the total numbers introduce an oversampling
at intermediate redshifts (e.g. those at $z\approx 0.09-0.1$). These
discrepancies are small enough that they do not produce any noticeable bias
in the most relevant statistical quantities that we have explored (see
next sections). We note that one could in principle compensate for
this by modifying the selection functions used in the method, for
which a mock galaxy survey will be needed for adequate calibration.
However, because in this paper focuses on the simplicity of the
method, in the following we do not introduce any such fine-tuning
compensation.

 \begin{figure*}
    
    \begin{minipage}{0.48\textwidth}
    \centering
    \includegraphics[width=1.01\textwidth]{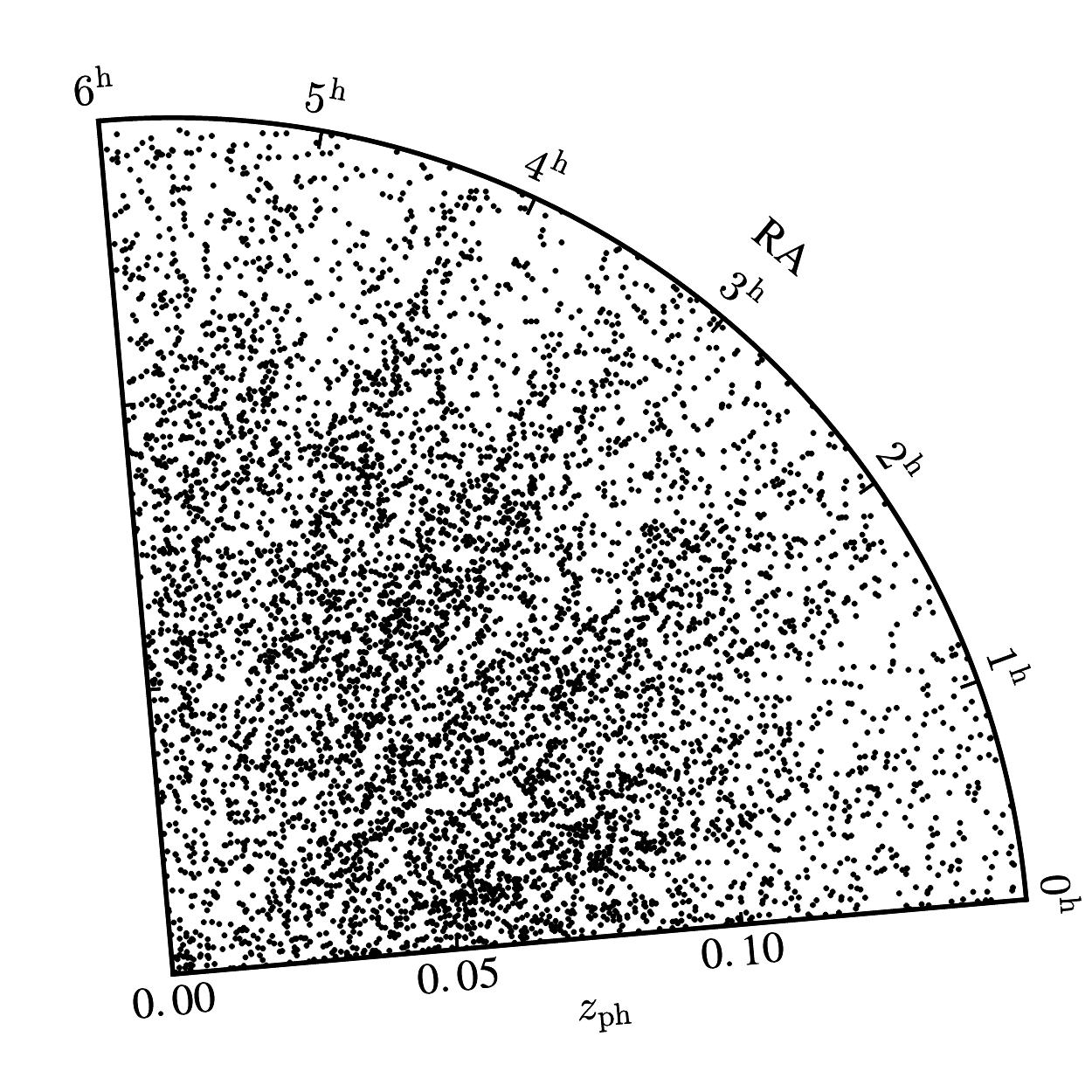}
    \end{minipage}
    \begin{minipage}{0.48\textwidth}
    \centering
    \includegraphics[width=1.01\textwidth]{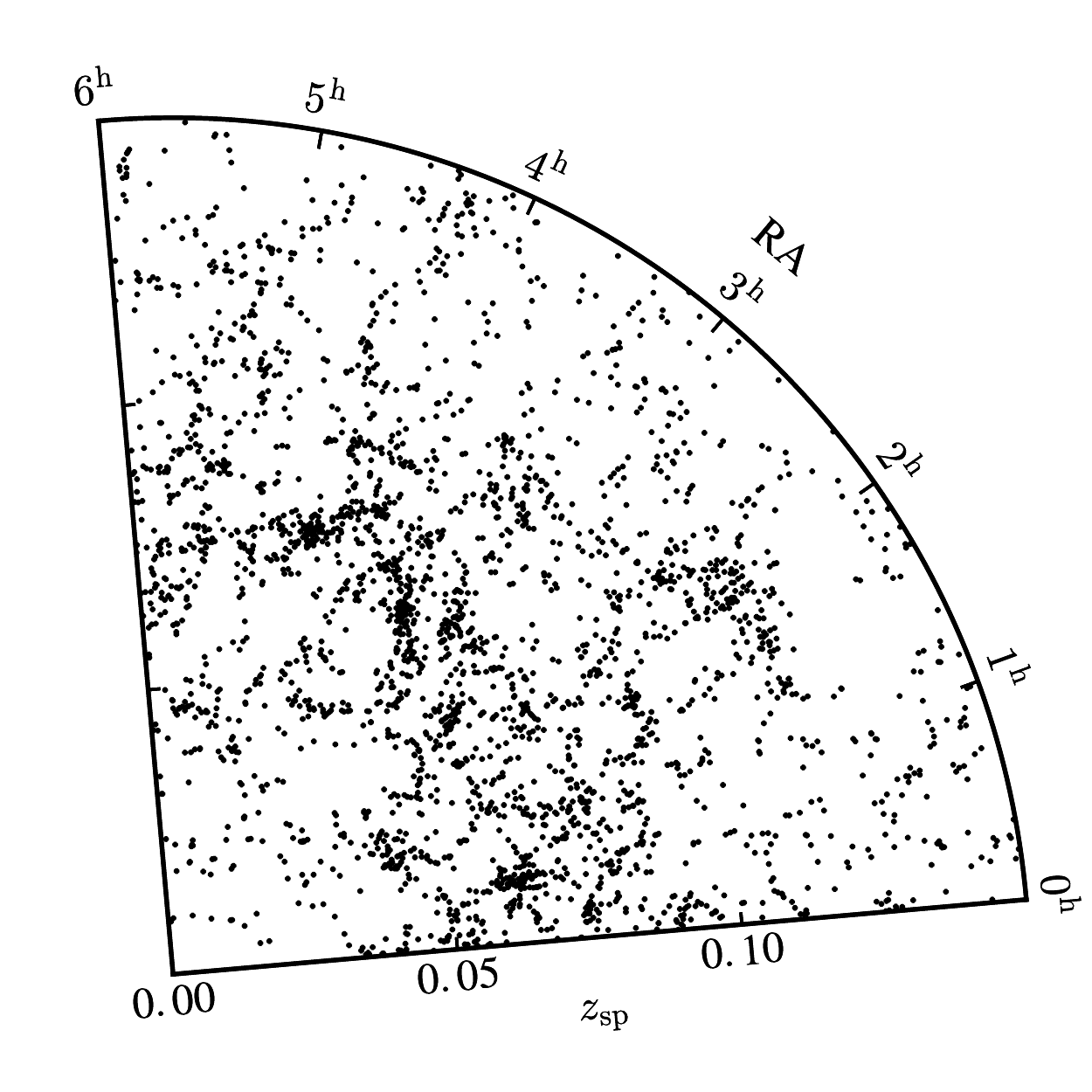}
    \end{minipage}
    \begin{minipage}{0.48\textwidth}
    \centering
    \includegraphics[width=1.01\textwidth]{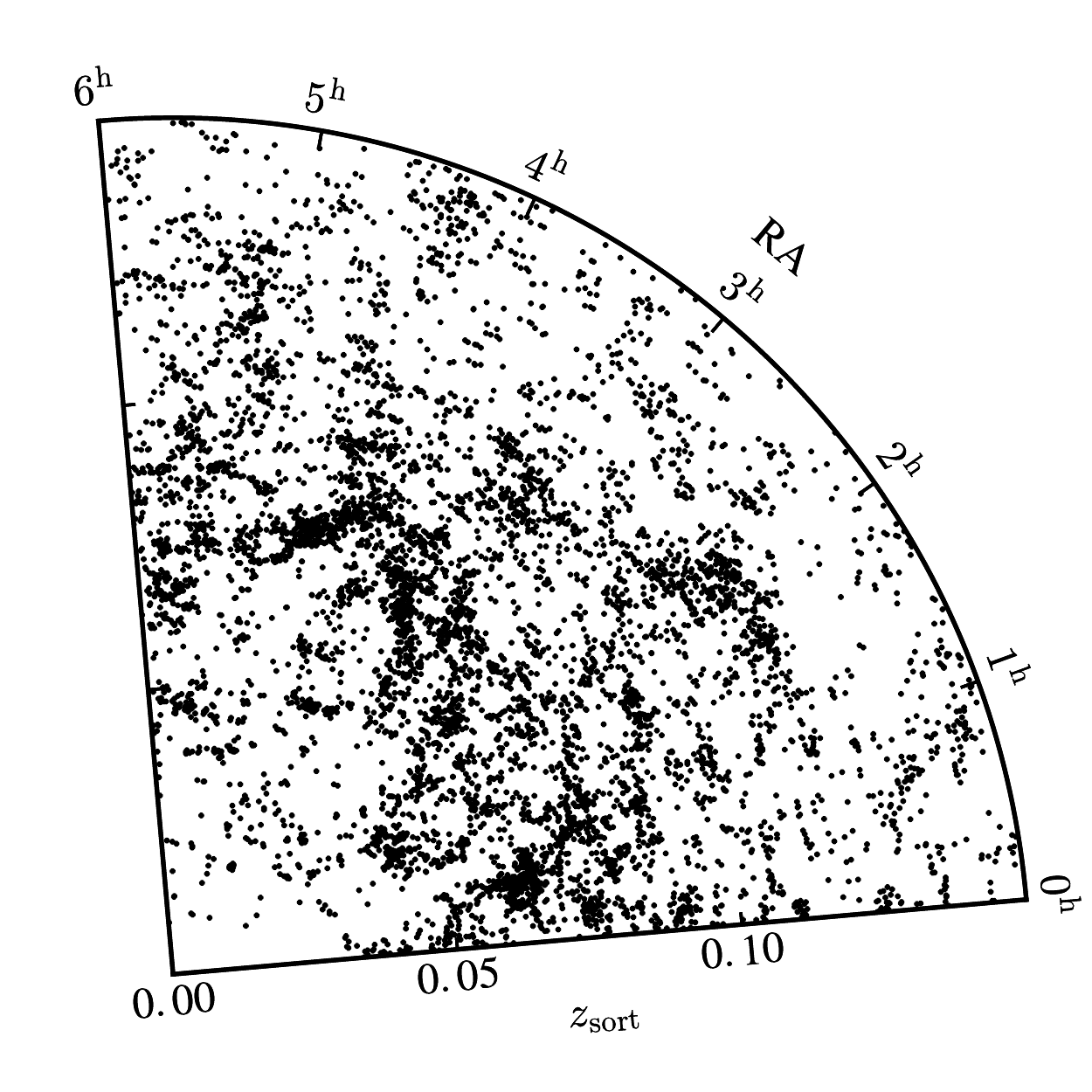}
    \end{minipage}
    \begin{minipage}{0.48\textwidth}
    \centering
    \includegraphics[width=1.01\textwidth]{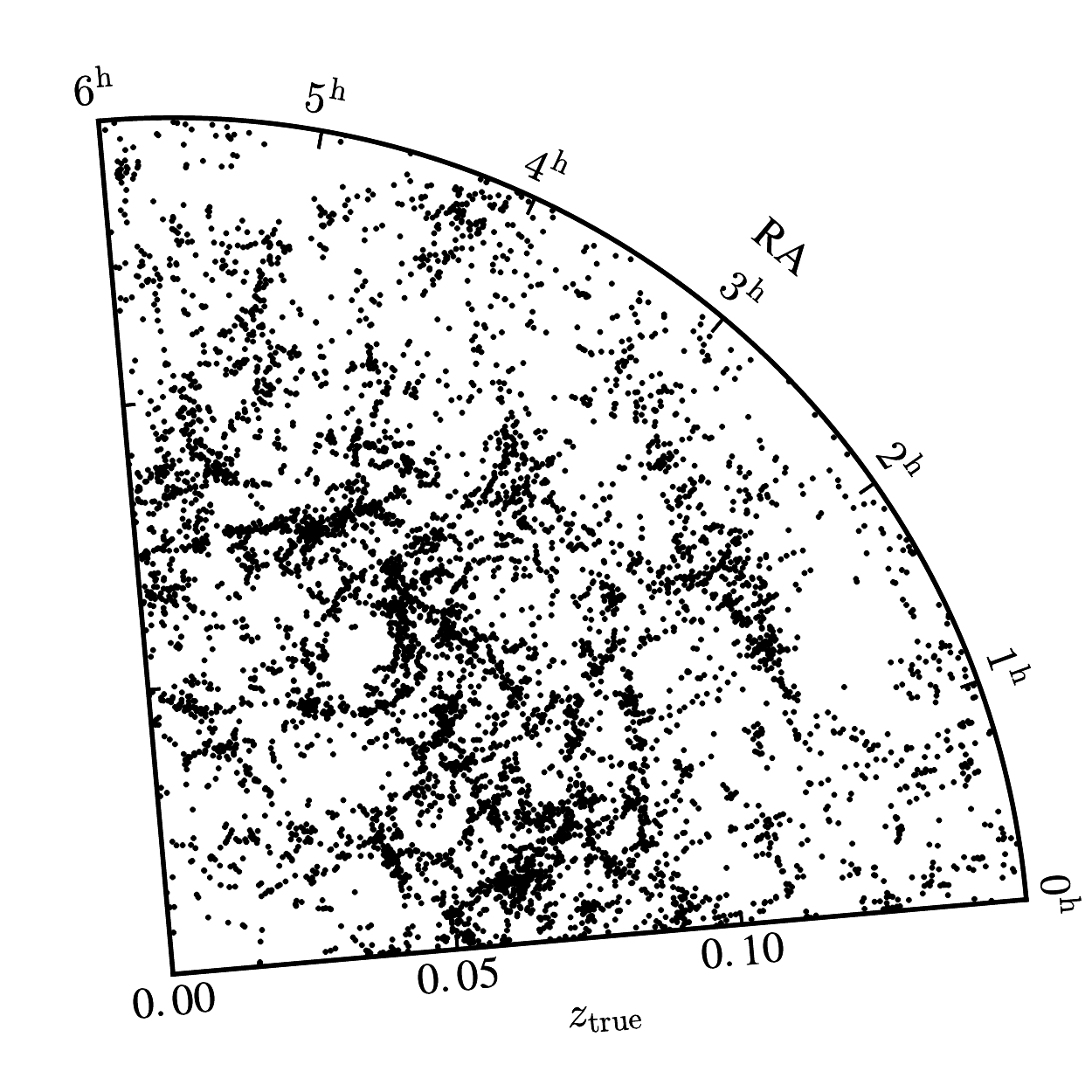}
    \end{minipage}
    
    \caption{A declination slice (arbitrary thickness of $5$
degrees) of the three-dimensional distribution of our mock galaxy
survey, using different redshift measurements: original
photometric ($z_{\rm ph}$; top left), original spectroscopic used
as reference ($z_{\rm sp}$; top right), \sort ($z_{\rm sort}$;
bottom left), and true ($z_{\rm true}$; bottom right). The radial
axes correspond to redshift direction while the azimuthal axes
correspond to right ascension of the Celestial Coordinates in
hours. See \Cref{sec:3d-dist} for further details.}

\label{fig:pie_diagrams}

\end{figure*}

\subsection{Improving the original redshift measurements}\label{sec:improving}

In order to test whether our \sort
redshift ($z_{\rm sort}$) has actually improved the redshift
estimation compared to the original photometric redshifts
($z_{\rm ph}$), we directly compare them with the underlying true
redshifts ($z_{\rm true}$). In \Cref{fig:2dhist_comparison} we
show two-dimensional normalized histograms for $z_{\rm ph}$ (left
panel, $y$-axis) and $z_{\rm sort}$ (right panel, $y$-axis) as a
function of $z_{\rm true}$ ($x$-axes). We observe that \sort
redshifts have a considerably reduced the scatter around the true
redshifts, at least for small differences. Although barely
appreciable in these plots however, we note that an underlying
broader scatter remains for larger redshift differences which
makes the whole distribution to have a comparable standard
deviation of $\sigma \approx 0.02$.

This behaviour is better illustrated in \Cref{fig:hist_comparison},
where we show the full redshift difference distributions for both
$z_{\rm ph}$ (red histogram) and $z_{\rm sort}$ (black histogram). By
construction, the original photometric sample has a Gaussian
uncertainty distribution with standard deviation $\sigma_{z_{\rm ph}}
\approx 0.02$. On the other hand, $z_{\rm sort}$ has a somewhat
symmetric (skewness of $\approx -0.3$) uncertainty distribution with a
much smaller full width at half maximum of FWHM\,$\approx 0.004$. This
is the peak that dominates the signal in the right panel of
\Cref{fig:2dhist_comparison}. If we model this narrow peak as a single
Gaussian, we obtain a standard deviation of $\sigma_{z_{\rm sort}}^{\rm
peak} \approx \frac{\rm FWHM}{2.4}\approx 0.0017$, which is about an order
of magnitude improvement with respect to the original photometric
uncertainty. At redshift differences $|\Delta z| \gtrsim 0.002$
however, the \sort redshift sample has a broader underlying
scatter, which makes the  overall distribution have a
dispersion comparable to that of the original photometric uncertainty
(i.e. $\sigma_{z_{\rm sort}} \approx 0.02$).

If both distributions have comparable standard deviations, do the
\sort redshifts represent a real {\it improvement} over the
original photometric redshifts? In order to answer this question,
one must carefully consider the meaning of the standard deviation
as a measure of statistical uncertainty. Indeed, it is worth
reminding the reader the standard deviation alone is not enough
to characterize an arbitrary PDF, even though in astronomy
is widely used as the main statistical uncertainty indicator
(partly because astronomers usually deal with Gaussian
distributions for which such an approach is justified). The
overall shape of the $\Delta z|_{\rm sort}$ distribution is not
Gaussian, as it has a much more peaked distribution (and consequently
more extended tails) for a similar mean ($\approx 0$) and similar
standard deviation ($\approx 0.02$). Indeed, based on its
kurtosis value of $\approx 5$, $\Delta z|_{\rm sort}$ resembles
more a Hyperbolic Secant instead. Here we argue that a more peaked
redshift uncertainty distribution around $0$ contains more
information than a less peaked one, and that having a
distribution like $\Delta z|_{\rm sort}$ is indeed an improvement
over $\Delta z|_{\rm ph}$.

This fact can be partially appreciated by our previous comparison of
the overall redshift distributions (see \Cref{fig:z_dists},
\Cref{sec:z_dists}). We can observe that the original photometric
sample not only loses information regarding the positions of the
large-scale structure (i.e. peaks and valleys) but also produces 
biases in the expected number of galaxies as a function of redshift.

Another way to look at this issue is by quantifying the $\Delta
z|_{\rm obs} \equiv z_{\rm obs} - z_{\rm true}$ as a function of
$z_{\rm obs}$. In \Cref{fig:2dhist_zdiff} we show the mean of $\Delta
z|_{\rm obs}$ as a function of $z_{\rm obs} = z_{\rm ph}$ (left panel)
and $z_{\rm obs} = z_{\rm sort}$ (right panel). We observe that indeed
the photometric sample is highly biased towards negative $\Delta z$
values at the lowest redshift bins and somewhat biased towards
positive $\Delta z$ values at the highest redshift bins. In contrast,
$z_{\rm sort}$ are virtually unbiased across the full redshift range
which is clearly an improvement.

In the following subsections we explore how \sort performs at
retrieving the three-dimensional large-scale structure distribution as
well as the two-point autocorrelation function.

\subsubsection{Three-dimensional spatial distributions}\label{sec:3d-dist}

\Cref{fig:pie_diagrams} shows a subvolume of the mock survey,
where the redshifts come from our different estimates: $z_{\rm
ph}$ (top left panel), $z_{\rm sp}$ (top right panel), $z_{\rm
sort}$ (bottom left panel) and $z_{\rm true}$ (bottom right
panel). Unsurprisingly, the original photometric sample does not
provide a good description of the cosmic web. On the other hand,
our \sort method is able to reproduce most of the significant
large-scale-structures in the volume, including voids, dense
filaments, groups and clusters (compare the two bottom panels to
each other). Part of this success is due to the chosen parameters
for applying the method (i.e. those relevant for the cosmic web
scales; see \Cref{sec:apply}) in combination with having a
representative reference sample (i.e., $z_{\rm sp}$; see the top
right panel of \Cref{fig:pie_diagrams}). The recovered
distribution is not perfect however and some noticeable artefacts
exist. For example, the $z_{\rm sort}$ cosmic walls are thicker
and the voids less empty than for $z_{\rm true}$. These should in
principle be overcome by fine-tuning of the parameters for
applying the method (specially $A$ and $\delta z$; see
\Cref{sec:apply}). Although the cosmic web seems to be reasonably
well recovered as a {\it whole}, we also emphasize that there
will be some individual galaxies assigned to the wrong structures
along the line-of-sight, i.e. those contributing to the tails of
the $\Delta z$ distribution (see \Cref{sec:improving}). Still,
the fact that $z_{\rm sort}$ provides a considerable fraction of
galaxies assigned to the {\it right} structures along the
line-of-sight (i.e. the narrow peak in $\Delta z$ distribution),
makes this \sort method promising for further studies aimed to
quantify the three-dimensional distribution of the cosmic web and
cosmological environment of galaxies using photometric redshifts.

\subsubsection{Two-point correlation functions}\label{sec:2pcf}

Although the two-point correlation function (2PCF) is not enough to
describe the full three-dimensional spatial distribution, it is a well
defined and simple statistical measurement that contains relevant
information regarding spatial clustering. As such, it has been applied
to a wide variety of extragalactic questions not just related to the
large-scale structure but also to galaxy evolution and cosmology.

\begin{figure}
    
    \begin{minipage}{0.48\textwidth}
    \centering
    \includegraphics[width=1\textwidth]{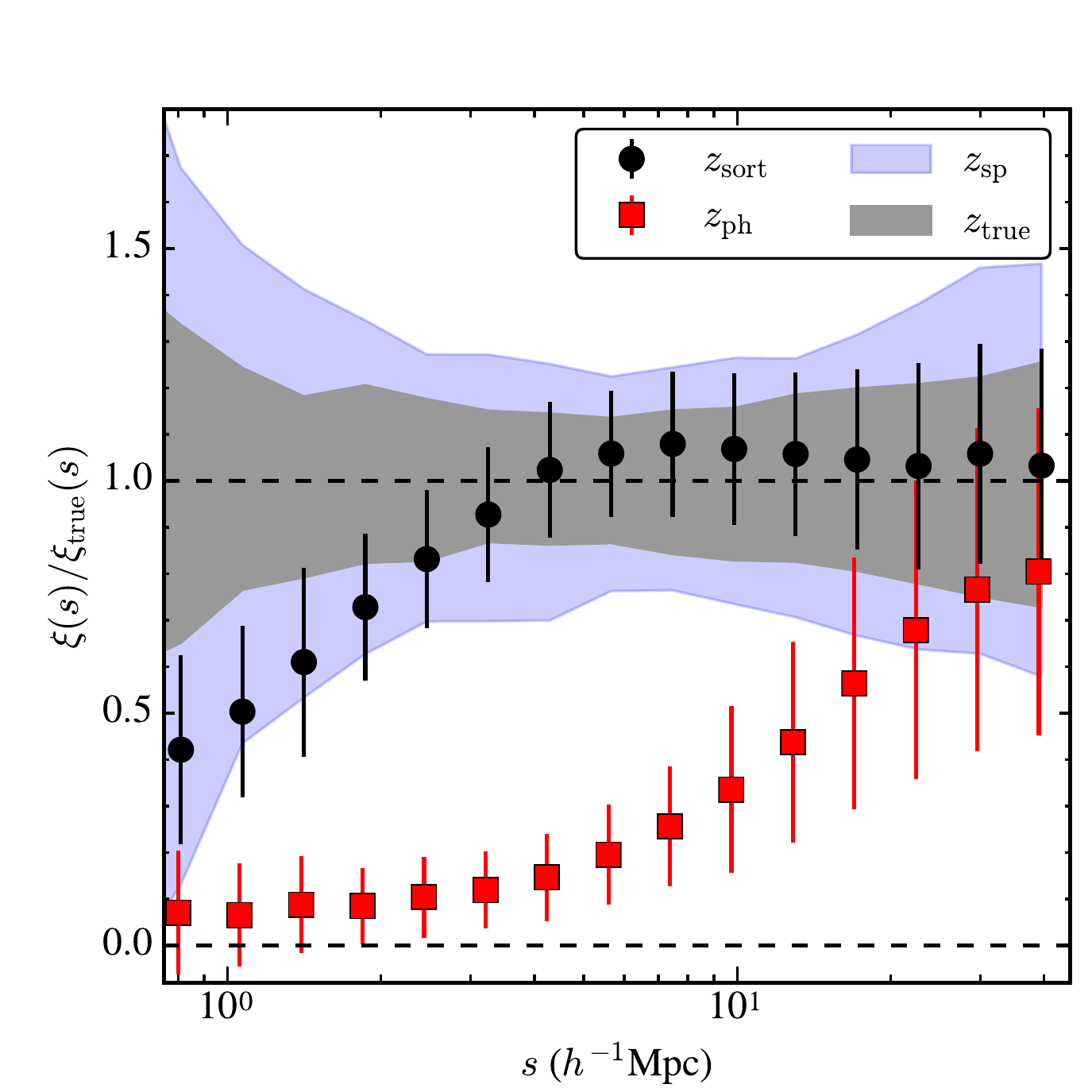}
    \end{minipage}
    
    \caption{The ratio between the measured redshift-space
two-point correlation function and its underlying true value,
$\xi(s)/\xi_{\rm true}(s)$, as a function of redshift space
distance $s$. The black circles correspond to using $z_{\rm
sort}$ and the red squares correspond to using $z_{\rm ph}$
(slightly offset in the $x$-axis for clarity). Uncertainties were
estimated from a bootstrap technique from $100$ realizations. The
grey shaded area corresponds to the intrinsic $1\sigma$
uncertainty limit due to sample variance (i.e. this is the
uncertainty assuming we knew the underlying true redshift for all
the galaxies in the original photometric sample). The light-blue
area corresponds to the $1\sigma$ uncertainty around the unbiased
$z_{\rm sp}$ measurement from the spectroscopic sample (i.e. the
remaining $30\%$ of galaxies used as reference). See
\Cref{sec:2pcf} for further details.}

\label{fig:xcorr_v1}

\end{figure}

\Cref{fig:xcorr_v1} shows the ratio between the measured redshift
space 2PCF and its true underlying value, $\xi_{\rm true}(s)$, as
a function of redshift space distance $s$. The black circles
correspond to using $z_{\rm sort}$ and the red squares correspond
to using $z_{\rm ph}$. Uncertainties were estimated with a
bootstrap technique from $100$ realizations. The grey shaded area
corresponds to the intrinsic $1\sigma$ uncertainty limit due to
sample variance (i.e. this is the uncertainty assuming we knew
the underlying true redshift for all the galaxies in the original
photometric sample). The light-blue area corresponds to the
$1\sigma$ uncertainty around the unbiased measurement using the
spectroscopic sample (i.e. the remaining $30\%$ of galaxies used
as reference; see \Cref{sec:mock}). This figure demonstrates that
$z_{\rm sort}$ is able to recover the true 2PCF on scales
$\gtrsim 4$\mpc in a somewhat unbiased manner at a higher
statistical precision than that of the reference spectroscopic
sample. At smaller than $\approx 4$\mpc scales however, $z_{\rm
sort}$ is not able to recover the 2PCF power; these scales are
comparable or smaller than that of the narrow peak recovered in
the redshift uncertainty distribution (see \Cref{sec:improving}).
On the other hand, the original photometric redshift are unable
to recover the 2PCF power at any scale, including those at
$\gtrsim 10-40$\mpc. We note that comparing in redshift space is
conservative because the redshift uncertainties introduce
distortions only along the line-of-sight. Our method preserves
the observed positions of objects in the sky, ensuring that the
angular correlations are not affected.

In view of these results, we argue that $z_{\rm sort}$ is not
just an improvement over $z_{\rm ph}$, but it also provides more
information than that contained in the reference sample used on
scales $\gtrsim 4$\mpc. {We note however that \sort was not
particularly optimized for measuring the two-point correlation
function, for which there may be other preferable methods
available.}

\section{Discussion}\label{sec:discussion}

\subsection{Where does the information come from?}\label{sec:information}

 \begin{figure*}
    
    \begin{minipage}{0.48\textwidth}
    \centering
    \includegraphics[width=1.01\textwidth]{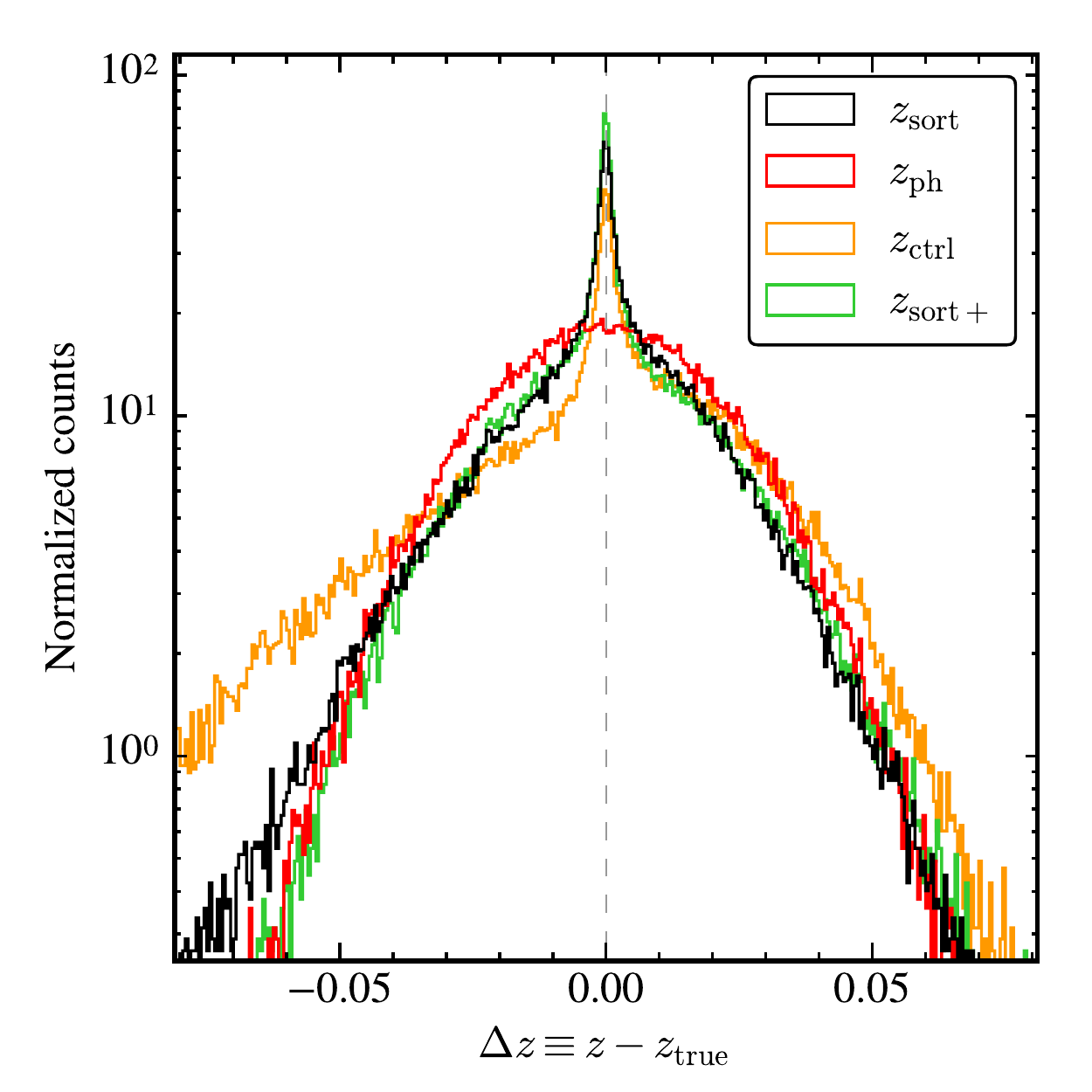}
    \end{minipage}
    \begin{minipage}{0.48\textwidth}
    \centering
    \includegraphics[width=1.01\textwidth]{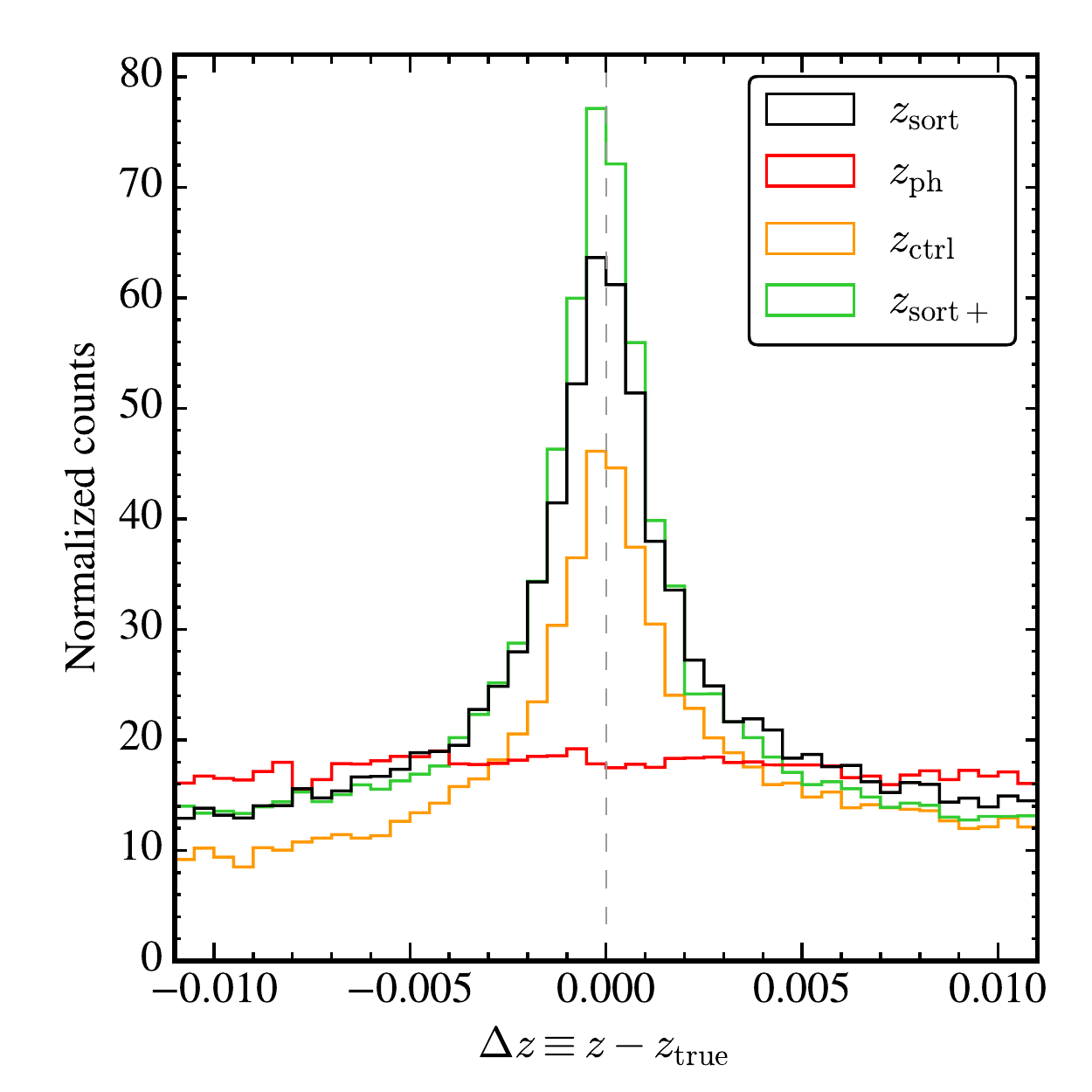}
    \end{minipage}
    
    \caption{Redshift difference $\Delta z \equiv z-z_{\rm true}$
for our different redshift measurements: $z_{\rm sort}$ (black),
$z_{\rm ph}$ (red), $z_{\rm ctrl}$ (orange), $z_{\rm sort+}$
(green). The left panel shows a log-linear scale, while the right
panel presents a zoom-in of the peak in a linear scale. See
\Cref{sec:information} for details.}.

\label{fig:information}

\end{figure*}

A relevant question to ask in the context of our method is where
the information comes from. Certainly, the use of a reference sample
must contribute a significant amount, but how much of an improvement
does the stochastic ordering matching scheme signify (see \Cref{sec:sort})? In
order to answer this question we have run a {\it control} method using
exactly the same algorithm presented in \Cref{sec:apply} but replacing
the fifth step simply with a random one-to-one assignment scheme
instead (i.e. no rank ordering was applied). We call this control
redshift assignment, $z_{\rm ctrl}$. In the following we will use the
$\Delta z \equiv z-z_{\rm true}$ normalized histograms as proxies of
the underlying redshift PDFs for our different methods. In principle,
a narrower and more peaked PDF contains more information than a
broader and less peaked one.\footnote{Note that in a hypothetical case
where $100\%$ of the information is recovered, the underlying PDF
will be a Dirac delta function; on the other hand, in a hypothetical
case where the information is completely lost, the underlying PDF will
be a uniform distribution.}

\Cref{fig:information} shows versions of \Cref{fig:hist_comparison}
including $z_{\rm ctrl}$ (orange histograms) and $z_{\rm ctrl+}$
(green histograms; described below); $z_{\rm sort}$ (black histograms)
and $z_{\rm ph}$ (red histograms) are the same as in
\Cref{fig:information}. The left panel shows $\Delta z$ histograms in
logarithmic scale, allowing us to emphasize the behaviour of the tails
of the distributions. The right panel shows $\Delta z$ histograms in
linear scale, with emphasis on the narrow peak. Comparing the orange
($z_{\rm ctrl}$) to the black ($z_{\rm sort}$) histograms we see that
the stochastic ordering provides extra information: the tails are more
symmetrical and more suppressed, and the peak of around $\Delta z = 0$
is higher.

From \Cref{fig:information} we can also see that $z_{\rm sort}$
is not necessarily the most optimal redshift estimation because
at the $|\Delta z|$ tails, there is a (small) contribution of
$z_{\rm sort}$ redshifts whose differences with respect to their
true values  are {\it larger} than what would be allowed by the
original photometric PDF, i.e. catastrophic redshift assignments.
We remind the reader that in obtaining $z_{\rm sort}$ we have so
far not used the information provided by the individual $z_{\rm
ph}$ uncertainties, but only their most likely redshift values.
(This was a deliberate choice in order emphasize the intrinsic
power of the stochastic order.) In order to take into account the
full individual $z_{\rm ph}$ PDFs, one can simply convolve these
with the distributions obtained at the end of the algorithm
presented in \Cref{sec:apply} to increase the quality of the
\sort. We call the redshifts from this optimal \sort method
$z_{\rm sort+}$, and its redshift uncertainty distribution is
shown as the green histograms in \Cref{fig:information}. We see
that indeed, $z_{\rm sort+}$ ensures a tail no larger than that
of the original $z_{\rm ph}$, and has a greater peak than that of
$z_{\rm sort}$.

We conclude that although a significant amount of information comes
from the use of the reference sample\footnote{Indeed, out method
cannot recover large-scale structures that are, for some reason, not
traced by the reference sample.}, the stochastic order matching scheme
does indeed add valuable extra information. In particular, \sort
significantly improves the quality of the recovered redshift PDF, as
it reduces the global asymmetry, suppresses the large tails, and makes
a more pronounced peak around $\Delta z = 0$. Moreover, an extra
enhancement can be easily achieved by using the information provided
by the full photometric PDFs rather than only the most likely value
alone.

 \begin{figure*}
    \begin{minipage}{0.48\textwidth}
    \centering
    \includegraphics[width=1.01\textwidth]{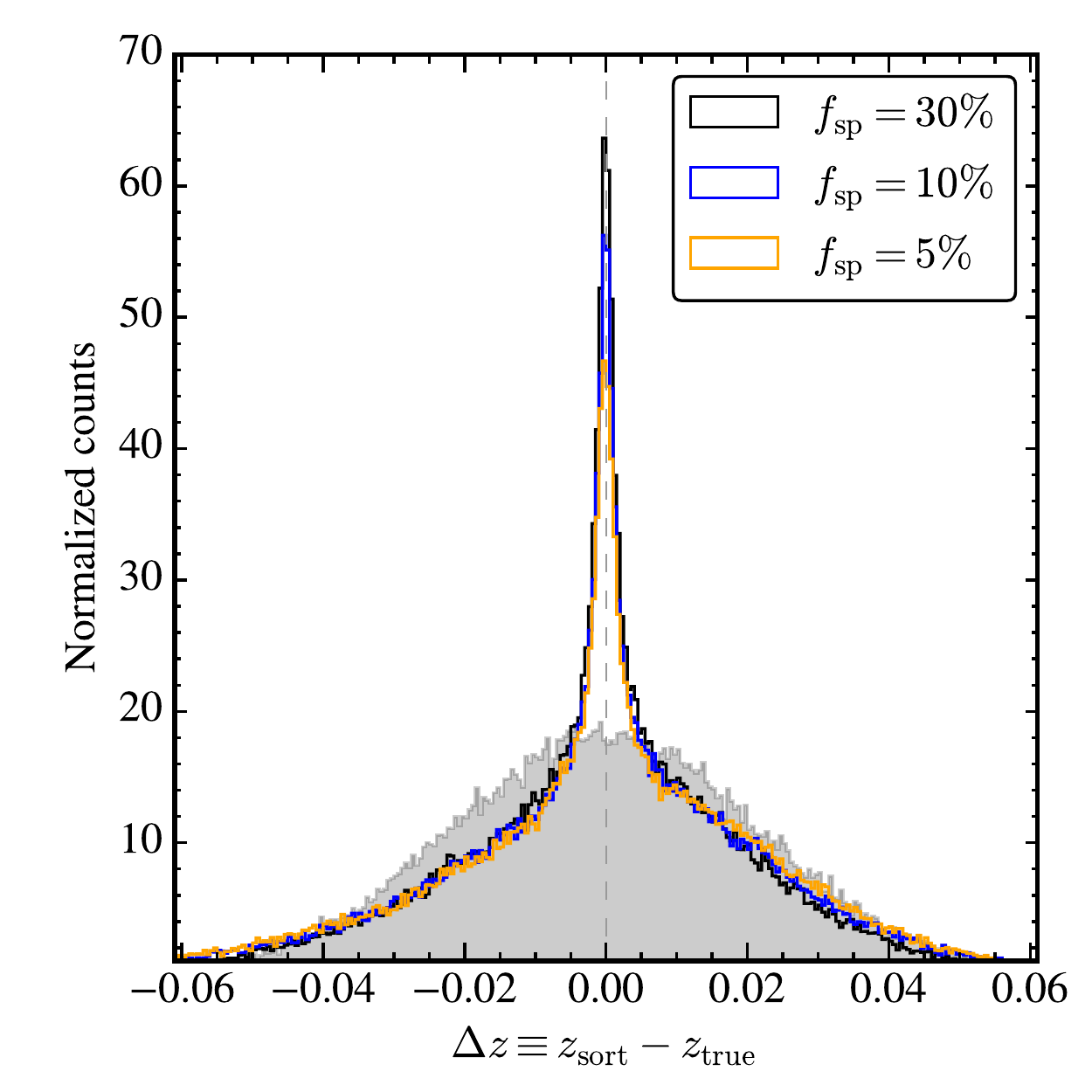}
    \end{minipage}
    \begin{minipage}{0.48\textwidth}
    \centering
    \includegraphics[width=1.01\textwidth]{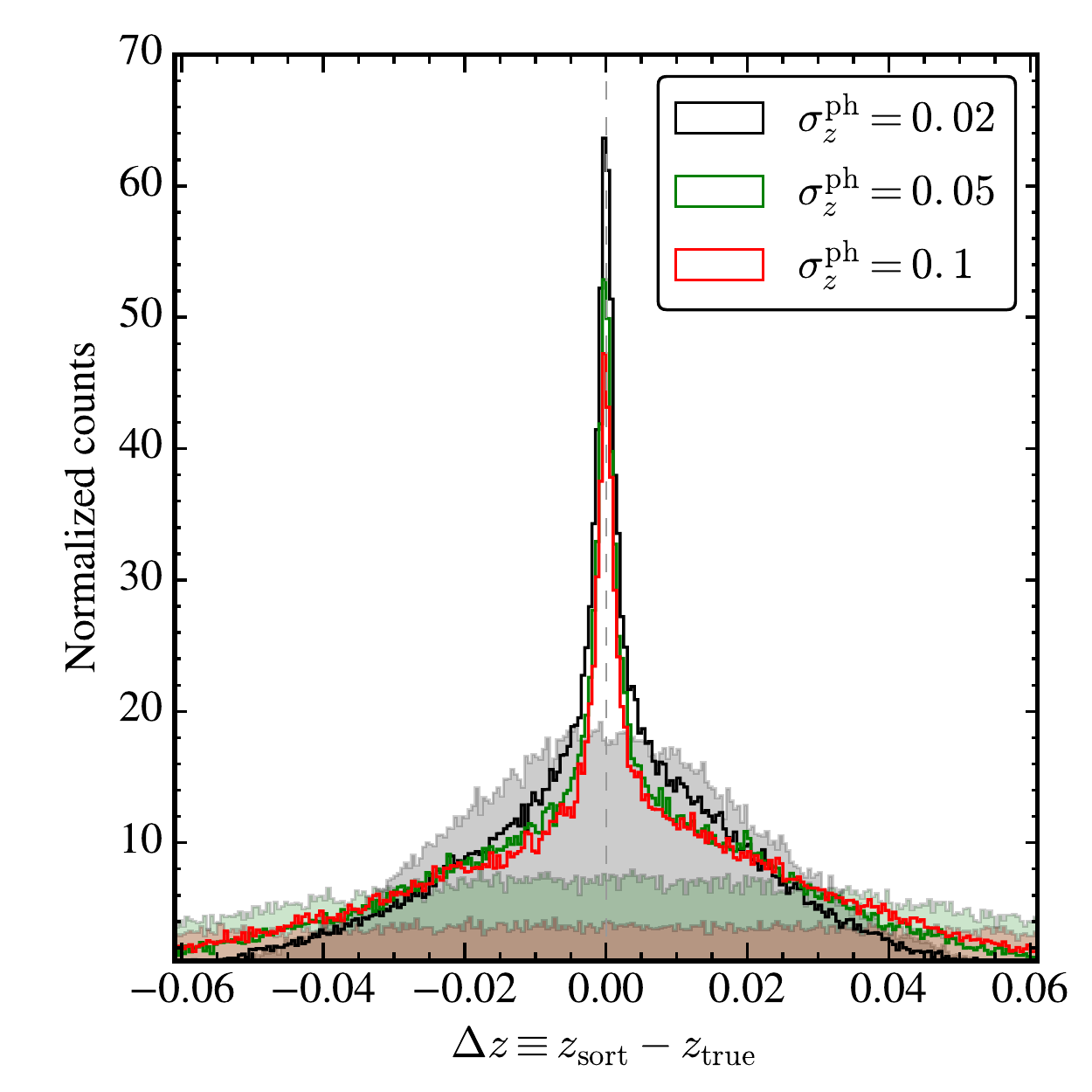}
    \end{minipage}
    
    \caption{{\it Left panel:} redshift difference $\Delta z \equiv
z_{\rm sort} - z_{\rm true}$ using different percentages of galaxies
as reference (spectroscopic): $f_{\rm sp}= 30\%$ (our fiducial value;
black histogram), $f_{\rm sp}= 10\%$ (blue histogram) and $f_{\rm sp}=
5\%$ (orange histogram). As reference, the original $\Delta z = z_{\rm
ph} - z_{\rm true}$ distribution is shown as a shaded grey. {\it Right
panel:} redshift difference $\Delta z \equiv z_{\rm sort} - z_{\rm
true}$ using different values for the original redshift uncertainties:
$\sigma_z^{\rm ph} = 0.02$ (our fiducial value; black histogram),
$\sigma_z^{\rm ph} = 0.05$ (green histogram) and $\sigma_z^{\rm ph} =
0.1$ (red histogram). As reference, the original $\Delta z = z_{\rm
ph} - z_{\rm true}$ distributions are shown as the shaded grey,
light-green  and light-red histograms, respectively. See
\Cref{sec:versatility} for further details.}

\label{fig:efficiency}

\end{figure*}

\subsection{Versatility and efficiency}\label{sec:versatility}

Perhaps one of the most attractive qualities of \sort is its
simplicity, which in turns makes it very versatile. Indeed, in its
general form the projected area $A$ can have any arbitrary shape and
size. Thus, the method can be applied to wide and narrow extragalactic
surveys alike (provided that suitable reference samples exist). We
also emphasize that \sort is intrinsically a non-parametric method, as
we are not assuming any functional form on the shape of the galaxy
distributions along individual lines-of-sight, nor on the luminosity
function of galaxies. Similarly, we are not applying convolutions nor
de-convolutions in order to infer the intrinsic underlying true
redshifts, which is advantageous in the sense that information is not
being deliberately lost.

Besides its versatility, \sort is also intrinsically efficient. The
most expensive requirement of the method is the use of a suitable
spectroscopic reference sample. However, considering that a
significant fraction of luminous galaxies reside in a small fraction
of the volume (i.e. luminous galaxies are biased tracers of the
underlying clustered matter distribution), then small reference
samples can still give a sensible mapping of the structures along a
given line-of-sight. If this is the case, then our proposed \sort
should still perform relatively well even using small reference
samples. In order to illustrate this point, in the left panel of
\Cref{fig:efficiency} we show normalized $\Delta z \equiv z_{\rm sort}
- z_{\rm true}$ for three \sort realizations (using the algorithm
presented in \Cref{sec:apply}), each one applied to mock galaxy
surveys as described in \Cref{sec:mock} having three different
percentages of reference spectroscopic galaxies: $f_{\rm sp}= 30\%$
(our fiducial value; black histogram), $f_{\rm sp}= 10\%$ (blue
histogram) and $f_{\rm sp}= 5\%$ (orange histogram). As expected, the
lower the fractions of galaxies used as reference, the poorer the
quality of the recovered $z_{\rm sort}$. Still, we see that even when
using a reference sample as small as $\sim 5\%$, \sort is able to
improve over the original photometric redshifts.

Moreover, as pointed out in \Cref{sec:sort}, if the individual
photometric redshift PDFs satisfy \Cref{eq:order} (i.e. stochastic
order), then the method should perform relatively well independently
of the actual value of the photometric redshift uncertainty. This is
illustrated in the right panel of \Cref{fig:efficiency}, where we show
normalized $\Delta z \equiv z_{\rm sort} - z_{\rm true}$ distributions
for three levels of photometric redshift uncertainty: $\sigma_z^{\rm
ph} = 0.02$ (black histogram; our fiducial value), $\sigma_z^{\rm ph}
= 0.05$ (green histogram) and $\sigma_z^{\rm ph} = 0.1$ (red
histogram). We observe that the narrow peak around $\Delta z = 0$ is
also well recovered using $\sigma_z^{\rm ph} = 0.05$ or $\sigma_z^{\rm
ph} = 0.1$. Although the quality of the recovered $z_{\rm sort}$ is
higher for lower $\sigma_z^{\rm ph}$, we emphasize that the {\it
relative} improvement over the original $z_{\rm ph}$ is actually
higher for larger $\sigma_z^{\rm ph}$. Indeed, in the later cases even
the standard deviation of $\Delta z$ is substantially reduced, from
the original values of $\sigma_z \approx 0.05$ and $\sigma_z \approx
0.1$ both to $\sigma_z \approx 0.03$.

We also consider \sort to be an efficient method in terms of
computational time. Indeed, the fact that \sort provides improved
cosmological redshifts for {\it samples} rather than individual
objects, can significantly reduce the cost of the method in CPU time.
Additionally, the most time consuming part of the algorithm, i.e. the
adaptive Monte Carlo realizations, can be easily parallelized as these
are all independent of each other.

\subsection{Applications}\label{sec:applications}  

{We consider that the main application for a method like
\sort would be to recover the distinctive features of the cosmic
web, i.e. estimating the locations of voids, filaments and walls.
These can then be used for studies of galaxy evolution as a
function of large-scale environment, by enabling  individual
galaxies to be located in their corresponding cosmic web
structures along the line-of-sight. Similarly, the \sort
redshifts can be used to estimate environmental density on scales
$\sim 4$\mpc, which correlates (directly or indirectly) with
several properties of dark matter halos \citep[e.g.][]{Lee2017}
or galaxies \citep[e.g.][]{Yan2013, Eardley2015}. This
application can also be extended to studies of the intergalactic
medium (IGM) in different cosmic environment traced by galaxies
\citep[e.g.][]{Penton2002, Stocke2007, Tejos2012, Tejos2016};
given that the IGM is usually detected through absorption lines
in the spectra of background QSOs, a method that could correctly
identify cosmic web structures in pencil-beam-like galaxy surveys
becomes particularly useful and efficient.

Another potential application may be on the characterization the
intrinsic properties of these large-scale structures, in
particular their shapes. Indeed, there is relevant cosmological
information imprinted in the size distribution of galaxy voids,
their relative underdensities, their density profiles, among
others \cite[e.g.][]{Sutter2014a,Sutter2014b,Cai2015,Hamaus2015,
Massara2015,Yang2015,Hamaus2016, Hamaus2017, Kovacs2017}. It is still to be seen whether \sort redshifts can
be used to improve results on these cosmological tests. }

\subsection{Limitations}\label{sec:limitations}

As any method, \sort does not lack limitations. The most obvious
one is the need of a reference sample to map out the real
structures along the line-of-sight. As shown in
\Cref{sec:information}, most of the information comes from the
reference sample. Then, if for some reason a structure is not
covered by the reference, \sort will not be able to recover such
structure. Similarly, there is also the intrinsic limit imposed
by the actual velocity dispersion of galaxies within the 
large-scale-structure of $\Delta v \approx 200$\,\kms, 
corresponding to redshift differences of $\delta z \approx 0.001$
(or $\sim 3$\mpc), which prevents \sort from reaching higher
precisions.

{An additional limitation comes from the actual selection
functions for both the photometric (uncertain) and the
spectroscopic (reference) samples, $S_{\rm ph}$ and $S_{\rm sp}$ 
(see \Cref{sec:general}), in the sense that relevant information is only obtained for the
redshifts range where both overlap. In the extreme case where
$S_{\rm ph}$ and $S_{\rm sp}$ are disjoint, our method cannot be
applied. Similarly, if the objects used as reference trace the
cosmic web {\it differently} than those in the uncertain sample (as it
may be the case for quasars versus dwarfs galaxies), then our
method will introduce a systematic error by wrongly assigning the
uncertain sample to match the cosmic web of the reference. Of
course, one can avoid this potential bias by making sure that the
reference sample is statistically relevant to that of the
uncertain one before applying \sort.}

Another intrinsic limitation of the method is that the recovered
mapping of structures is only robust for the {\it full} ensemble, but
not necessarily for individual objects. This comes from the fact that
we cannot distinguish which individual galaxies contribute to the peak
of the $\Delta z$ distribution, and which ones contribute to the
tails. This means that despite having a sensible reconstruction of the
three-dimensional distributions (see \Cref{sec:3d-dist}) and/or the
two-point correlation function on scales $\gtrsim 4$\mpc, there will
be galaxies assigned to the wrong structures along the line-of-sight.
Although $z_{\rm sort}$ could be in principle used to define
large-scale environments (voids, filaments, clusters), still a more
precise redshift estimation will be needed to place individual
galaxies into these environments.

We also note that the efficiency of \sort is limited by the
choice of the projected area $A$ to define sub-volumes. Let us
consider two extreme cases as examples. When $A \to 0$ the method
will be applied to individual galaxies rather than the ensemble,
reducing its efficiency (1) because we loose the information
provided by the angular correlations, and (2) because we need a
comparatively larger fraction of galaxies in the reference sample
(more expensive).\footnote{Indeed, for the case $A=0$ we may need
galaxies right {\it on top} of the original targets which is of
course physically prohibiting.} On the other hand, when $A \to
4\pi$ the method will be applied to the whole sky, reducing its
efficiency because the information provided by the angular
correlations as a function of redshift will be washed out
(although it may still work for very local cosmological
structures). For simplicity, in this paper we have used a
circular $A$ of radius $1$\,degree chosen because it corresponds
to $\approx 5-10$\mpc at the redshifts where most of our galaxies
reside (see \Cref{sec:apply}). In principle however, there should
exist an optimal $A$ that maximizes the amount of information
recovered by \sort. This optimal $A$ has to be small enough to
resolve the angular correlations introduced by galaxy clustering
as a function of redshift, and large enough to include a minimum
number of reference redshifts ensuring a sensible mapping of the
structures. Thus, this optimal $A$ has to be a function of
redshift $A(z)$ and also a function of the survey design. {For instance, given the behaviour of transverse distance as a
function of redshift, we expect that for $z>1$, an optimal $A$
containing transverse scales up to $\sim 5-10$\mpc could be as
small as a tenth of a degree (i.e. few arcmins), making \sort
particularly promising for future generations of deep
pencil-beam-like surveys.} Obtaining a general optimization of
$A$ is beyond the scope of this paper. Surveys such as COSMOS and
CANDELS would be ideal to put constraints on the optimal $A(z)$
function and test the limitations of \sort.

\subsection{Comparison to previous work and complementarity}

In this section we provide a comparison with two recent relevant
published methods, namely the PhotoWeb \citep{Aragon-Calvo2015}
and the clustering-based redshift estimation \citep{Menard2013}.

{\it Comparison to PhotoWeb} Our method has many similarities with
the PhotoWeb proposed by \citet{Aragon-Calvo2015}. Both methods
rely on the presence of an statistically relevant reference
sample tracing an underlying `cosmic web' of structures along the
line-of-sight, and both methods use this reference sample in
order to improve photometric redshifts. However, an important
difference between SORT and PhotoWeb is that the latter acts on
individual galaxies rather than an ensemble. This is both
advantageous and disadvantageous. The advantage is that PhotoWeb
can reconstruct the underlying density field in a given
line-of-sight at a much better resolution than that of SORT; the
disadvantage is that PhotoWeb does not necessarily preserve
$dN/dz(z)$ as it tends to over-represent galaxy over-densities
and under-represent the galaxy under-densities. In other words,
their voids are too empty and their walls are too populated. This
is because the improved PhotoWeb redshift estimation comes from
multiplying different PDFs (i.e. density, geometrical and
photometric, see their equation 1) of individual galaxies
independently of the recovered redshifts of the others. On the
other hand, \sort starts from the condition that the $dN/dz$ of
the reference sample has to be preserved, and only uses the
photometric individual PDFs to establish their final rank order
in redshift.

Regarding the $\Delta z \equiv z-z_{\rm true}$ recovered
distributions, both techniques give similar shapes (see their
figure 9). Although \citet{Aragon-Calvo2015} have modelled it as
the combination of two Gaussians, here we have argued that based
on its kurtosis value, a Hyperbolic Secant may be more
appropriate description (see \Cref{sec:improving}). Moreover,
they have used the width of the narrow peak to claim a sub-megaparsec
precision (i.e. redshift uncertainty as low as $\approx 0.0007$).
However, because there is no way to determine  which individual
galaxies contribute to the narrow peak as opposed to the broader
$\Delta z$ distribution, such a low uncertainty estimation is way
too optimistic.

{\it Comparison to the clustering-based redshift estimation}
Another relevant recent work to compare with is that of
\citet{Menard2013}. {In contrast} to \sort or PhotoWeb, the
clustering-based method relies on a reference sample (e.g.
spectroscopic) {but it only requires the existence of a
non-zero  two-point correlation function, which is a weaker
constraint that requiring a well defined `cosmic web'}. The
clustering-based method is fully independent from the original
photometric redshift estimation, as it only uses the information
given by the {\it spatial} clustering of the sources. {In this
manner, the resulting clustering-based redshift PDF can be
multiplied with the independent photometric redshift PDF to
obtain an improved cosmological redshift for individual
sources.}\footnote{{See also \citet{Lee2016} for another
clustering-based method aimed at directly obtaining cosmological
redshifts for individual sources.}}  The essence of their
method is to divide the survey volume into multiple narrow
redshift slices, and for a given slice to identify the subsample
of galaxies that maximise the projected angular correlation to
that of the respective reference sample while making sure the
$dN/dz$ is also preserved. This is in contrast to our \sort
method, where we divide the survey volume into multiple
pencil-beam like surveys and impose the photometric samples to
match the $dN/dz$ of the respective reference sample in each of
those sub-volumes, while also preserving their rank order in
redshift. 

Because of its intrinsic simplicity and versatility (see
\Cref{sec:versatility}), we envision that a method like \sort
could be integrated into these more complex and sophisticated
algorithms aimed at improving cosmological redshifts.

\subsection{Future prospects}

The obvious next step is to apply \sort to real data (e.g. SDSS),
for which an accounting of the more complex selection function
and individual photometric redshift PDFs must be taken into
account. In such a case, the method should be calibrated against
a realistic mock survey for optimization.

Of particular interest for the upcoming large photometric surveys
(e.g. LSST), is to explore the efficiency of the method when
using much fainter galaxies, and whether or not one can
effectively use brighter (cheaper) galaxies as reference to
recover the structures along a given line-of-sight. Similarly,
one can also produce forecasts on how well the large-scale
structures (voids, {walls,} filaments, clusters) can be
recovered using \sort, and the impact of such reconstruction on
weak-lensing mapping, the recovery of the baryonic acoustic
oscillations (BAO), etc.

There is also some room for improvement of the method in terms of
using other galaxy properties beyond position alone, such as
star-formation activity, colour, shapes, etc. Moreover, our
current implementation does not explicitly take into account the
clustering information {\it within} the projected area $A$.
Indeed, by assuming a functional form of the spatial clustering
of galaxies one could assign join probabilities for galaxies to
be at a similar redshifts if they appear highly clustered within
$A$; such information can help to better constrain the original
individual PDFs, hence improving the overall performance of
\sort. {Although we expect our method to apply mostly to
field galaxies, for wider applications it will be necessary to
compensate for large galaxy redshift distortions introduced by
galaxy clusters (i.e. the so-called fingers-of-god),  which could
be done using standard techniques \citep[e.g.][]{Tegmark2004,
Aragon-Calvo2015}.}

In a more general context, we believe that the main strengths of
\sort are its simplicity and its versatility; hence, it would
make more sense to use it in combination with some other more
sophisticated method aimed at improving cosmological redshifts
\citep[e.g.][]{Menard2013, Aragon-Calvo2015}.

\section{Summary and conclusions}\label{sec:summary}

In this paper we have presented a simple, efficient and robust
method to improve cosmological redshifts with uncertainties
larger than the `cosmic web' scales ($\gtrsim 3$\mpc).   The
method is based on the presence of a reference sample for which a
precise redshift number distribution ($dN/dz$) can be obtained
for different pencil-beam-like sub-volumes within the original
survey. For each sub-volume we then impose: 

\begin{itemize}
\item (i) that the redshift number distribution of the uncertain
redshift measurement matches the reference $dN/dz$ corrected by their
selection functions; and 

\item (ii) the rank order in redshift of the original ensemble of
uncertain redshift measurements is preserved.  \end{itemize}

\noindent The latter step is motivated by the fact that random
variables drawn from Gaussian probability density functions
(PDFs) of different means and arbitrarily large standard
deviations satisfy stochastic ordering. We then repeat this
simple algorithm for multiple arbitrary pencil-beam-like
overlapping sub-volumes; in this manner, each uncertain
measurement has multiple (non-independent) ``recovered''
redshifts which can be used to estimate a new redshift PDF.
Assuming that redshift measurements can be treated as random
variables, we thus refer to this method as the Stochastic Order
Redshift Technique (\sort).

We have used a state-of-the art $N$-body simulation to test the
performance of \sort under simple assumptions and found that it
can improve the quality of cosmological redshifts in an efficient
manner. Particularly, \sort redshifts ($z_{\rm sort}$) are able
to recover the distinctive features of the so-called `cosmic web'
and can provide unbiased measurement of the two-point correlation
function on scales $\gtrsim 4$\mpc. Based on these results, we
have argued that \sort provides more information than that of the
reference sample alone.

Given its simplicity, we envision that a method like \sort can be
incorporated into more sophisticated algorithms aimed to exploit
the full potential of large extragalactic photometric surveys.
Indeed, most of the state-of-the-art photometric (and clustering
based) redshift algorithms reach individual statistical
precisions of $\sigma_z\approx 0.02-0.05$ (Gaussian). In this
paper we have shown that such precision can be further improved
applying an extremely simple, efficient and robust method that
relies on stochastic order.

\section*{Acknowledgements} 

{We thank the anonymous referees for their constructive
criticism and comments that improved the quality of this manuscript.}  We thank
Miguel Aragon-Calvo, Brice M\'enard, Bahram Mobasher and J.
Xavier Prochaska for useful comments and discussions. N.T.
acknowledges support from {\it CONICYT PAI/82140055} and from the
IMPS Fellowship\footnote{\url{http://imps.ucolick.org}} at
University of California, Santa Cruz, where part of this work was
conducted. A.R.P. was partially supported by a UC-MEXUS
Fellowship.  J.R.P. acknowledges support from grant
HST-GO-12060.12-A-004. We thank the Leibniz-Rechenzentrum (LRZ)
in Munich where the MultiDark-Planck simulations were run on the
SuperMUC supercomputer. We also thank the Leibniz Institute for
Astrophysics Potsdam (AIP) and the Spanish MultiDark Consolider
project for supporting the MultiDark and CosmoSim databases. We
thank contributors to SciPy\footnote{\url{http://www.scipy.org}},
Matplotlib\footnote{\url{http://www.matplotlib.sourceforge.net}},
Astropy \footnote{\url{http://www.astropy.org}
\citep{AstropyCollaboration2013}}, the Python programming
language\footnote{\url{http://www.python.org}}, and the free and
open-source community.

\bibliographystyle{mn2e_trunc8}
\bibliography{/media/ntejos/disk1/lit/bib/IGM}

\bsp
\label{lastpage}
\end{document}